\documentclass[journal]{IEEEtran}
\usepackage[utf8]{inputenc}
\usepackage[T1]{fontenc}
\usepackage{amsmath,amsfonts, bm}
\usepackage{algorithmic}
\usepackage{algorithm}
\usepackage{array}
\usepackage[caption=false]{subfig}
\usepackage{textcomp}
\usepackage{stfloats}
\usepackage{url}
\usepackage{verbatim}
\usepackage{graphicx}

\usepackage[normalem]{ulem}

\usepackage{stmaryrd}
\usepackage{cite}
\usepackage{color}
\usepackage{hyperref}
\usepackage{cleveref}       
\usepackage{xr}             

\usepackage{booktabs} 
\usepackage{array}    
\usepackage{caption}  
\usepackage[table]{xcolor} 
\usepackage{multirow}  
\usepackage{rotating}  

\definecolor{dkred}{rgb}{0.8, 0.0, 0.0}
\definecolor{dkgreen}{rgb}{0.0, 0.5, 0.0}
\definecolor{dkblue}{rgb}{0.0, 0.0, 0.7}
\definecolor{dkorange}{rgb}{1.0, 0.3, 0.0}

\newcommand{\QUESTION}[1]{\textbf{ \color{dkblue} QUESTION:} {\color{dkblue} #1}}
\newcommand{\ANS}[1]{\textbf{ \color{dkorange} ANSWER:} {\color{dkorange} #1}}
\renewcommand{\QUESTION}[1]{}
\renewcommand{\ANS}[1]{}

\begin{document}

\title{The Marginal Importance of Known Distortions and Alignment in CASSI systems}
\author{Léo Paillet, Antoine Rouxel, Hervé Carfantan, Simon Lacroix and Antoine Monmayrant

\thanks{This work was supported by the CNRS and the Université de Toulouse. Léo Paillet is with LAAS-CNRS, Toulouse, France and with IRAP, Toulouse, France (e-mail: leo.paillet@laas.fr). Antoine Rouxel, Simon Lacroix and Antoine Monmayrant are with LAAS-CNRS, Toulouse, CNRS, France (e-mails: {\em firstname.name}@laas.fr). Hervé Carfantan is with IRAP, Toulouse, France (e-mail:Herve.Carfantan@irap.omp.eu). This paper has supplementary downloadable material available at http://ieeexplore.ieee.org., provided by the author. The material includes a supplementary document describing some charasteristics of our systems and more detailed reconstruction results. This material is 8.5MB in size.}
}

%
\markboth{IEEE TRANSACTIONS ON COMPUTATIONAL IMAGING}{Paillet \MakeLowercase{\textit{et al.}}: The Marginal Importance of Known Distortions and Alignment in CASSI systems}

\maketitle

\begin{abstract}

   Coded Aperture Snapshot Spectral Imagers (CASSI) are designed to acquire 2D coded acquisitions of hyperspectral scenes (HSS).
	Associated with reconstruction algorithms, they allow to analyze the HSS using a small amount of data.
	Various configurations of such imagers exist, with different distortions. 
	In this article we show that reconstruction quality is relatively insensitive to these distortions, if known and correctly modeled. 
   To this end, we introduce a differentiable ray-tracing-based model that incorporates aberrations and distortions to render coded hyperspectral acquisitions using CASSI.
   Such simulated acquisitions are used to train state-of-the-art hyperspectral cube reconstruction algorithms.
   We also adapted these algorithms through the use of a ray-tracing-based mapping function which accounts for aberrations and distortions.
   We evaluated four comparable CASSI systems with varying degree of optical aberrations and misalignments, using five state-of-the-art hyperspectral cube reconstruction algorithms.
   Our analyses show that if known and properly modeled, the effects of geometric distortions of the system and misalignment of the dispersive elements have a marginal impact on the overall reconstruction quality.
   Therefore, relaxing traditional constraints on measurement conformity and fidelity to the scene enables the development of novel imaging instruments, guided by performance metrics applied to the design or the processing of acquisitions. 
   By providing a complete framework for design, simulation and evaluation, this work contributes to the optimization and exploration of new  CASSI systems and more generally to the computational imaging community.

\end{abstract}


\section{Introduction}
\IEEEPARstart{T}{raditional} hyperspectral imagers render three-dimensional data cubes by scanning the scenes along a spectral or spatial dimension~\cite{Garini2006}.
This leads to two main drawbacks: the transmission and processing of a large quantity of data, and a limitation to static scenes.
Snapshot hyperspectral coded aperture imaging systems~\cite{Yuan2021, Wagadarikar:08, Arce2014, Cao2016, Gehm:07} rely on compressed sensing theory~\cite{Donoho2006} to address these shortcomings.
Specifically, CASSI systems~\cite{Wagadarikar:08, Gehm:07} reduce the redundancy present in hyperspectral scenes (HSSs) by performing encoding of the contained information.
Out of the many different implementations of CASSI, the two most common are DD-CASSI~\cite{Gehm:07} and SD-CASSI~\cite{Wagadarikar:08, Wagadarikar:09}.
Thanks to the double disperser, DD-CASSI induces mainly spectral encoding, while SD-CASSI induces both spectral and spatial encoding.
These systems use a coded aperture (or ``mask'') and dispersive elements to spectrally and spatially modulate the HSS.
The coded aperture can be optimized for specific applications~\cite{Wang2018}, but can also be used as a random sampler~\cite{Wagadarikar:09, Ardi2018}.

A primary goal of coded hyperspectral imaging is the reconstruction of the full hyperspectral cube (HSC), enabling further processing of the three-dimensional data cube.
Historically, this reconstruction from coded acquisitions predominantly relied on model-based methods~\cite{Kittle:10, DeSCI, GapTV, Ardi2018, Fu2016}.
While these methods offer insight into the reconstruction process, they suffer from long reconstruction times and suboptimal reconstruction quality.
Recently, deep learning approaches have provided fast reconstruction times and high-quality results, albeit with less interpretability~\cite{Huang2022, Wang2018, Miao2019, TSANet}. 
Transformers~\cite{AttentionIs} have also achieved superior results by leveraging non-local spatial relationships between pixels~\cite{MST, CST},
resulting in better exploitation of spatial and spectral interrelationships in HSSs.

However, state-of-the-art data-driven reconstruction algorithms~\cite{DGSMP,MST,DAUHST,RDLUF,PADUT}, also integrate a simplified representation of the propagation model, with a linear spreading on the detector and without any optical distortion or misalignment.
This approximation is best suited for grating-based CASSI\cite{Lin2014,McGregor2015,Saita:22,Lin:14}, but most CASSI systems are using prism-based dispersive element that offer higher efficiency for large spectral ranges.
This approximation is also well suited for most CASSI systems that employ a double-Amici prism as the dispersive element~\cite{Wagadarikar:09, Song:22, Wang2015, Wang2015opt, Xiong2017, Feng2014}.
Compared to single prisms, the advantage of double-Amici prism assemblies is twofold: they allow for a direct-view geometry and do not exhibit anamorphic or optical distortions.
Still, double-Amici prisms require longer manufacturing times, are more expensive, and misalignments must still be addressed when working with prototypes.
Alternatively, using a single prism as the dispersive element results in a peculiar arrangement of information in the captured images, which can be properly exploited for reconstruction if the optical model is precise and accurate enough.

In this article, we evaluate the impact of known distortions and misalignments arising from the dispersive element on the coded information, when the optical model is properly considered.
For this matter, we need a CASSI simulator accounting for both distortions and the spatially and spectrally Point Spread Function (PSF).
Some image formation models~\cite{Song:22} include PSF but they depend on calibration data, necessitating the assembly and calibration of a prototype.
Recently, we proposed a chief-ray-based propagation model~\cite{Rouxel2024} for dimensioning and optical distortions estimation, but it does not account for optical aberrations and cannot be used for Monte-Carlo rendering or PSF estimation.
Recent works in computational imaging utilize differentiable ray-tracing to enable end-to-end optical design~\cite{Sun2021, Tseng2021, dO}, in particular DiffOptics (dO)~\cite{dO}.
However, to the best of our knowledge, these works only address optical systems with axial symmetry thus cannot be used for coded aperture hyperspectral systems.

We propose an extension of dO to more general optical systems -- in particular breaking axial symmetry -- which thus enables modeling of optical behavior and precise rendering of coded hyperspectral acquisitions.

Although it does not simulate the PSFs, we can harness the ray-traced spot diagrams to render coded acquisitions encompassing distortions and aberrations.
By leveraging ray tracing, we create a spectral and spatial mapping between the object and image planes of SD-CASSI systems.
This mapping is then introduced in the reconstruction algorithms. 
This approach facilitates seamless processing across different CASSI systems, regardless of distortions or misalignments. 
We evaluate four distinct SD-CASSI configurations, constructed with either a double-Amici prism assembly or a single prism. To assess the impact of misalignments, two of these configurations are deliberately misaligned. 
The reconstructions from these four configurations are compared using five reconstruction algorithms and standard evaluation metrics.

Our results demonstrate that distortions and misalignments have a marginal impact on the information encoded in SD-CASSI acquisitions, and hence on the reconstruction quality, provided they are known and properly accounted for in the acquisition rendering and incorporated into the reconstruction algorithms.
This finding implies that the choice and fine-tuning of the reconstruction algorithm are more critical than the specific optical system.

The contributions of our work are:
\begin{itemize}
	\item We implement coded aperture hyperspectral optical systems within a differentiable ray-tracing framework, enabling the rendering of distorted and aberrated images from backward ray-tracing and facilitating the end-to-end design of such systems before prototyping.
	\item We devise a spatio-spectral mapping based on a ray-traced propagation model and introduce it, as a spatio-spectral prior information, in the state-of-the-art reconstruction algorithms. 
	This approach leverages two-dimensional measurements from the  system and the associated rendering model.
	\item We highlight the marginal importance of distortions and misalignments in CASSI systems for acquisition processing purposes, provided the reconstruction process incorporates a propagation model including known distortions.
\end{itemize}
The paper organization is described as follows.
Section~\ref{sec:design_simulation} starts by explaining the design of a double-Amici prism assembly with the same angular spectral spreading than an off-the-shelf single prism, but none of its optical distortions.
Two comparable SD-CASSI systems are then simulated --one based on the single prism, one on the double-Amici prism assembly--, and two configurations are considered for each : one with a perfectly aligned prism and one with a purposefully misaligned prism.
These four configurations allow for the validation of our differentiable ray-tracer by comparing the obtained spot diagrams with a well-known ray-tracer (Zemax).
The section ends with the rendering process of CASSI acquisitions for each configuration through our ray tracer.
Afterwards, Section~\ref{sec:performance_evaluation} details the processing of coded acquisitions to reconstruct HSCs with algorithms taking into account the propagation model.
Reconstruction quality is finally compared across all four configurations, for five state-of-the-art algorithms, demonstrating that the performance metrics are similar across configurations for a given algorithm.
When misalignment and distortions are unknown, performances deteriorate depending on the configuration.

All the source code about our modified version of dO, the adapted networks, and how to generate the figures is available at \url{https://github.com/lpaillet-laas/DiffCassiSim}.

\section{Optical design and rendering}
\label{sec:design_simulation}

Given that CASSI systems performances cannot be fairly assessed and compared using simple propagation models~\cite{Ardi2018,Diaz2021,Arguello2013}, which overlook the geometric distortions caused by dispersive elements as well as misalignment, we have developed a differentiable ray-tracing-based renderer based on dO, that takes into account distortions, misalignments and aberrations.

\begin{figure}[!t]
	\centering
	\includegraphics[width=0.95\columnwidth]{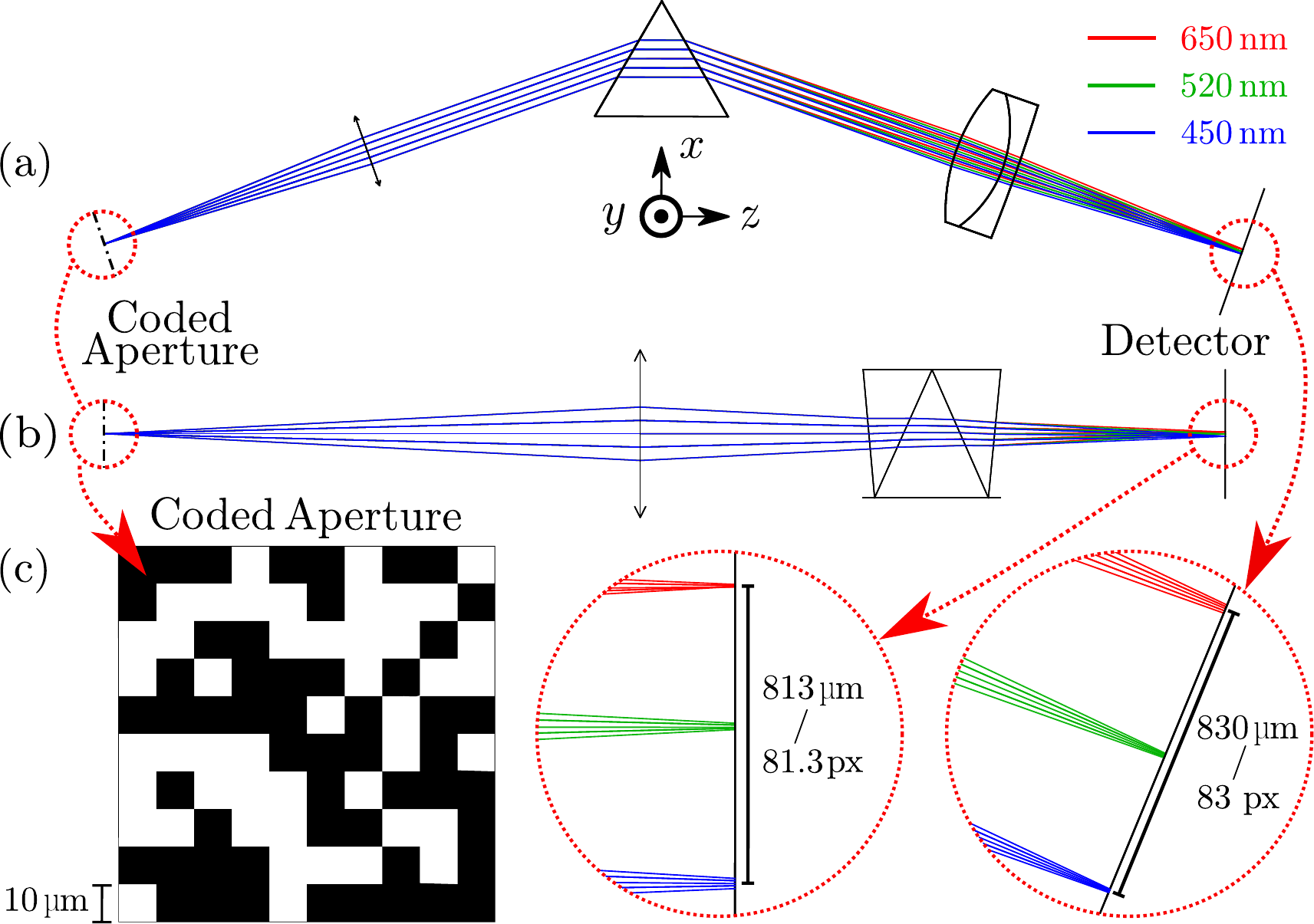}
	\caption{
      Top-view of the systems layout for the (SP) configuration (a) and the (AP) configuration (b).
      An example of a coded aperture and a zoom on the detector planes (c).
      The imaging lens is not pictured here, we consider that the scene is perfectly imaged on the coded aperture's plane.
   }
	\label{fig:system_layout}
\end{figure}

We applied it to generate  rendering of the coded acquisitions of two single disperser (SD) CASSI systems: one system using a single prism as the dispersive element, referred as (SP), and one with a double-Amici prism assembly, referred as (AP), shown in Figure\,\ref{fig:system_layout}. The two other configurations are duplicates for which the dispersive element has been purposefully misaligned by 5° around the $x$-axis with respect to the reference frames of Figure\,\ref{fig:system_layout}, and respectively referred as (mSP) and (mAP).
We chose an exaggerated misalignment, far larger than what an experimenter should achieve, in order to consider the worst-case scenario and prove that even a large misalignment can lead to satisfactory processing and hence be considered marginal.

In this section, we delve into:
\begin{itemize}
	\item The design process of a custom double-Amici prism to minimize optical distortions and achieve direct-view.
	\item The implementation of the four optical configurations within dO, highlighting how different configurations are simulated.
	\item The methodology used for generating ray-traced coded hyperspectral acquisitions.
\end{itemize}

\subsection{Case Study: SD-CASSI}
\label{subsec:systems_settings}

The (SP) system illustrated in Figure\,\ref{fig:system_layout}-(a) contains a first lens to collimate the light onto the prism and a second one to image the light onto the detector.
For the (AP) system, as for most recent Amici-based systems~\cite{Wagadarikar:09, Wang2015}, a single relay lens is used in a \(2f-2f\) configuration, with the dispersive element positioned between the lens and the detector, as shown in Figure\,\ref{fig:system_layout}-(b).

To ensure a fair comparison between both systems, it is crucial to dimension the dispersive elements so that they exhibit comparable spectral dispersion. 
Our analysis is conducted on systems with a detector comprising $512\times512$ pixels with a 10-µm pitch resulting in a $\simeq 5 \times 5$\,mm$^2$ field of view, a spectral range of $[450- 650]$\,nm centered at $520$\,nm, and a numerical aperture of $0.05$. 
The systems utilize lenses with a focal length of $f = 50$\,mm.
The targeted angular spectral spreading for these systems is $\Delta_0 = 0.95$°, resulting in a spatio-spectral spreading $S$ on the detector of $S \simeq 830\,$µm (83 pixels) for the central point of the field of view.
Since the magnification of such systems almost equals one, we also consider a coded aperture with a 10\,µm-pixel-pitch.
These parameters are selected based on the following criteria: the expected number of resolved pixels across the field of view, the consistency with the spectral range and resolution of the KAIST dataset~\cite{Choi:17}, and the compatibility with standard off-the-shelf components.

\subsection{Design of a Double-Amici Prism for Fair Comparison}
\label{subsec:prism_design}

\begin{figure}[!t]
	\centering
	\includegraphics[width=0.95\columnwidth]{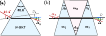}
	\caption{
      Optical designs of dispersive elements in our imaging system.
      (a) A commercially available N-BK7 equilateral prism, aligned in for minimum deviation $D_0$.
      (b) A custom-designed double-Amici prism optimized to minimize distortions and achieve direct-view geometry.
      Both designs exhibit comparable angular spectral spreading $\Delta$ to ensure a fair comparison of system performance.}
	\label{fig:prisms}
\end{figure}

Our first goal is to design a double-Amici prism that replicates the angular spectral dispersion $\Delta_0=0.95$° of a standard N-BK7 equilateral prism at minimum deviation $D_0$ illustrated in Figure\,\ref{fig:prisms}-(a), while minimizing geometric distortions and ensuring a direct-view configuration.

To design the prism, we employ the chief-ray-based dimensioning tool SIMCA, described in~\cite{Paillet2024,Rouxel2023}, which solves a gradient-based optimization problem incorporating both optical and system-related parameters. 
Although dO is more powerful for differentiable optical design, we chose SIMCA as it was specifically developed for CASSI systems, thus offering more straightforward and faster optimization.

\subsubsection{Prism Parametrization}

We optimize the following five parameters, illustrated in Figure\,\ref{fig:prisms}-(b):
$\alpha_c$ the angle of incidence on the double-Amici prism assembly,
$A_1$ and $A_2$ the apex angles of the first and second prisms, respectively,
and $m_1$ and $m_2$ the glass materials for the two prisms, selected from the Schott catalog.
Since the glass materials are discrete and not differentiable by default, we model the dispersion curve of each glass using two continuous sub-parameters: the refractive index at the "d" Fraunhofer line and the Abbe number, following the methodology described in~\cite{cote2023differentiable,Sun2021}. 
During optimization, we treat these sub-parameters as continuous variables. 
After optimization, we select the glass materials from the catalog that are closest to the optimized sub-parameter values.

\subsubsection{Loss Functions}

To optimize the prism design, we first aim to match the angular spatio-spectral dispersion of the base design $\Delta_0$. 
We then minimize the optical distortions across the field of view and for the whole spectral range.
We finally enforce an easy-to-align and compact system by minimizing the angular deviation $D$ and the thickness of the prism.

The \textit{spectral dispersion loss} $\mathcal{L}_{\Delta}$ is defined as the squared difference between the base spectral dispersion $\Delta_0$ and the dispersion $\Delta$ calculated for the current prism design. 
We compute $\Delta$ as the absolute difference between the output angles after the prism for the shortest and longest wavelengths at the center of the field-of-view, respectively denoted as $\Delta_{\lambda_{min}}$ and $\Delta_{\lambda_{max}}$:
\begin{equation}
\mathcal{L}_{\Delta} = \left( \Delta_0 - |\Delta_{\lambda_{max}} - \Delta_{\lambda_{min}}| \right)^2
\end{equation}

The \textit{distortion loss} $\mathcal{L}_{\varepsilon}$ quantifies the geometric distortions introduced by the prisms.
We calculate a distance tensor $\bm{\varepsilon}$ which measures how each imaged point of the scene is displaced due to distortion.
In practice, we measure the Euclidean distance between the distorted coordinates $(X_{\text{ds}}, Y_{\text{ds}})$ and the ideal coordinates $(X_{\text{id}}, Y_{\text{id}})$:
\begin{equation}
\bm{\varepsilon}_{i,j,k} = \sqrt{\left(X_{\text{ds}}^{i,j,k} - X_{\text{id}}^{i,j,k}\right)^2 + \left(Y_{\text{ds}}^{i,j,k} - Y_{\text{id}}^{i,j,k}\right)^2}
\end{equation}
Here, $\left(X_{\text{id}}^{i,j,k}, Y_{\text{id}}^{i,j,k}\right)$ are the coordinates of the ideal (undistorted) image grid points as described in~\cite{Rouxel2024}. 
We then define the distortion loss as the square of the maximum value in this distance tensor:
\begin{equation}
\mathcal{L}_{\varepsilon} = \left( \max_{i,j,k} \bm{\varepsilon}_{i,j,k} \right)^2
\end{equation}
The minimization of $\mathcal{L}_{\varepsilon}$ reduces the maximum geometric distortion across all points in the image grid.

The \textit{deviation loss} $\mathcal{L}_D$ accounts for the total angular deviation induced by the prism configuration. 
The deviation is computed based on the chief-ray angles $\alpha_c$ and $\alpha_c^{\text{out}}$ (incidence angle and output angle of the chief-ray at the central wavelength) and the apex angles of the prisms. 
The deviation loss is then given by the squared total deviation:
\begin{equation}
\mathcal{L}_{D} = \left( \alpha_c + \alpha_c^{\text{out}} + 2A_1 - A_2 \right)^2
\end{equation}
where $A_i$ are the apex angles of the prisms.

The \textit{thickness loss} $\mathcal{L}_{t}$ approximates the physical thickness of the double-Amici prism. It is proportional to the sum of the squared apex angles:
\begin{equation}
\mathcal{L}_{t} = 2A_1^2 + A_2^2
\end{equation}
This loss helps to minimize the overall size and weight of the optical system.

The \textit{glass distance loss} ($\mathcal{L}_{\text{g}}$) measures the squared difference between the refractive index ($n_d$) and Abbe number ($v_d$) of the selected glass materials and the closest available materials in the Schott catalog.
This loss ensures that the materials chosen during the optimization process are realistic and match closely with available catalog materials.
For a double-prism system, the glass distance loss is defined as:
\begin{equation}
\mathcal{L}_{\text{g}} = \gamma_g \cdot \min_{\text{glass}} \left( \left( \frac{n_d - n_d^{\text{catalog}}}{\Delta n_d} \right)^2 + \left( \frac{v_d - v_d^{\text{catalog}}}{\Delta v_d} \right)^2 \right)
\end{equation}
where $\Delta n_d$ and $\Delta v_d$ represent the ranges of refractive index and Abbe number in the Schott catalog. 
$\gamma_g$ is a weight parameter, currently set to $10^4$, enabling comparable values between losses inside the global composite loss.
This ensures that the chosen materials are practical and manufacturable.

Finally, the \textit{total internal reflection loss} ($\mathcal{L}_{\text{R}}$) penalizes designs that approach the critical angle for total internal reflection.
Although this metric can be further refined, it is currently defined as :
\begin{equation}
   \mathcal{L}_{\text{R}} =\log\left (1 + \exp(a-D_{\text{TIR}})\right )^b
\end{equation}
where $D_{\text{TIR}}$ is the angular distance from which total internal reflection would occur and $a$ and $b$ are tunable parameters (currently set to $5$ and $6$, respectively).

\subsubsection{Design Optimization}

The goal is to find the set of design parameters $\theta = \{\alpha_c, A_1, A_2, m_1, m_2\}$
that minimize the total loss $\mathcal{L}(\theta)$ defined as a linear combination of the 6 loss terms:
\begin{equation}
\mathcal{L} = w_{\Delta} \mathcal{L}_{\Delta} + w_{\varepsilon} \mathcal{L}_{\varepsilon} + w_D \mathcal{L}_{D} + w_t \mathcal{L}_{t} + w_g \mathcal{L}_{g} + w_R \mathcal{L}_{R}
\label{equ:loss} 
\end{equation}
$\mathcal{L}$ encapsulates both optical and system-related objectives, allowing for a comprehensive optimization of the prism design.
The weights $w$ in the loss expression \eqref{equ:loss} balance the contributions of each loss term according to their importance in the design objectives, they have been empirically chosen to guide the optimization.
Specifically, the values used in the optimization were $(w_{\Delta}, w_{\varepsilon}, w_D, w_t, w_{\text{g}}, w_{\text{R}}) = (1, 1, 25, 5, \text{iteration step}, 10)$, with all distances in mm and angles in mrad.
It is important to note that the choice of weights and loss functions should be tailored according to the specific design requirements.
Although not utilized in this study, exhaustive grid searches or Monte Carlo-based methods could be necessary for determining optimal weights and loss functions tailored to more demanding design constraints.
\subsubsection{Results}
\begin{figure}[!h]
	\centering
	\includegraphics[width=0.95\columnwidth]{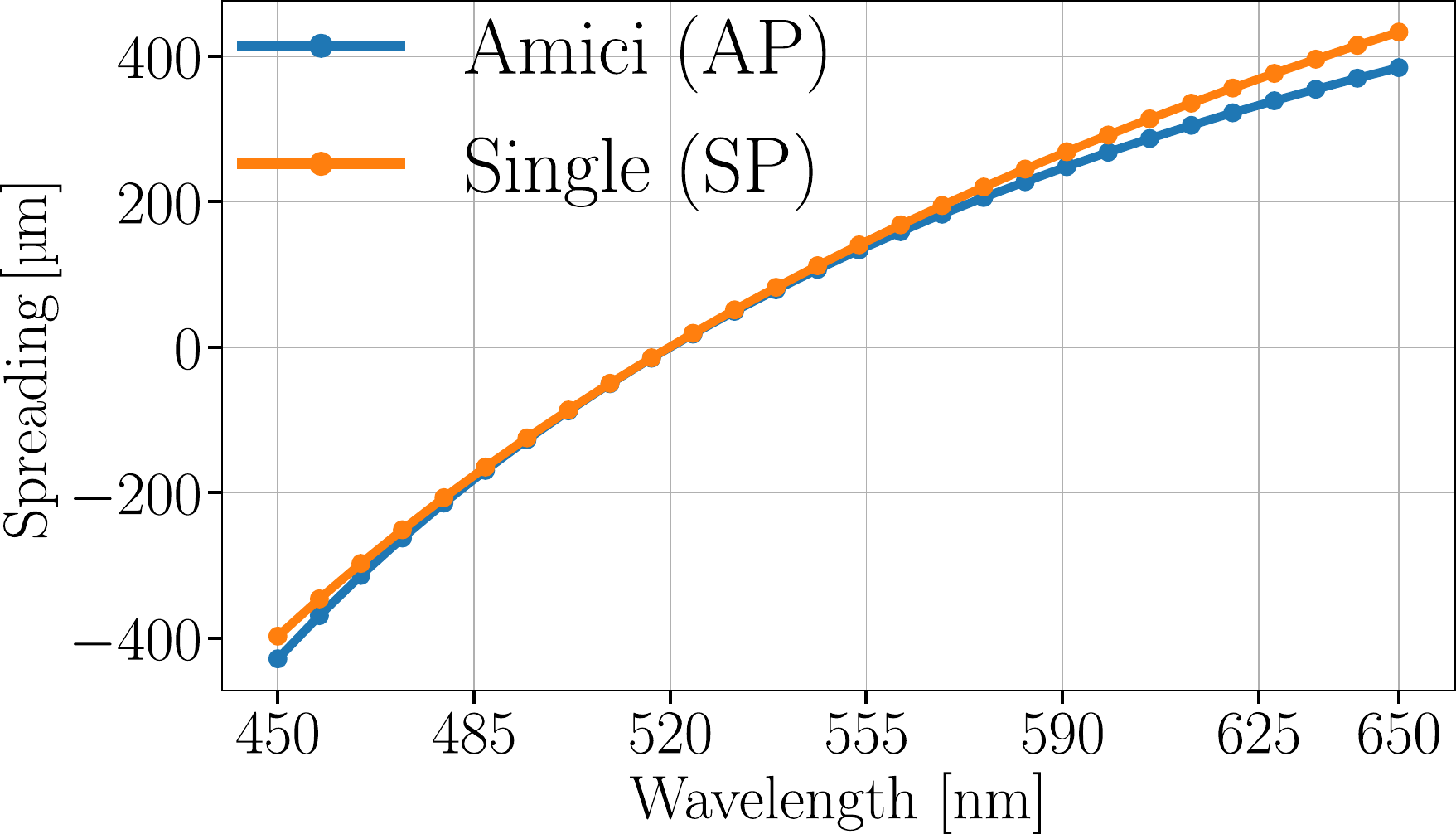}
      \caption{Comparison of the spatial dispersion at the detector (spreading) for (AP) (blue) and (SP) (orange).
      The dots correspond to 28 considered wavelengths over our spectral range.}
	\label{fig:dispersion_curves}
\end{figure}

Using the Adam optimizer~\cite{Kingma2014} and starting from the Amici parameters described in~\cite{Wagadarikar:09}, we design a double-Amici prism assembly that meets our spectral dispersion requirement $\Delta_0$ of 0.95° at the detector plane within $\pm2\%$.
The final materials for the Amici prism are N-SK2 and SK10 for \(m_1\) and \(m_2\), respectively. The apex angles are \(A_1 = 29.2\)° and \(A_2 = 47.9\)°, with the incident chief-ray arriving at the prism at \(\alpha_c = 5.1\)°.
Figure\,\ref{fig:dispersion_curves} presents the resulting spatial dispersion in the detector plane for the (AP) and the (SP).
This spreading corresponds to the wavelength-dependent spatial shift of the image of the central point.
As one can see, the curves have slightly different nonlinearities, up to 7.9\,pixel difference to the fitted linear spreading for (SP) and 11\,pixel for (AP) (see Supplementary Figure~\ref{supp-fig:nonlinear}).
Yet the overall spreading ranges on the detector are similar: 830\,µm for (SP) and 813\,µm for (AP).
Both systems thus exhibit comparable spectral dispersion.

\begin{figure}[h!]
	\centering
	\includegraphics[width=0.95\columnwidth]{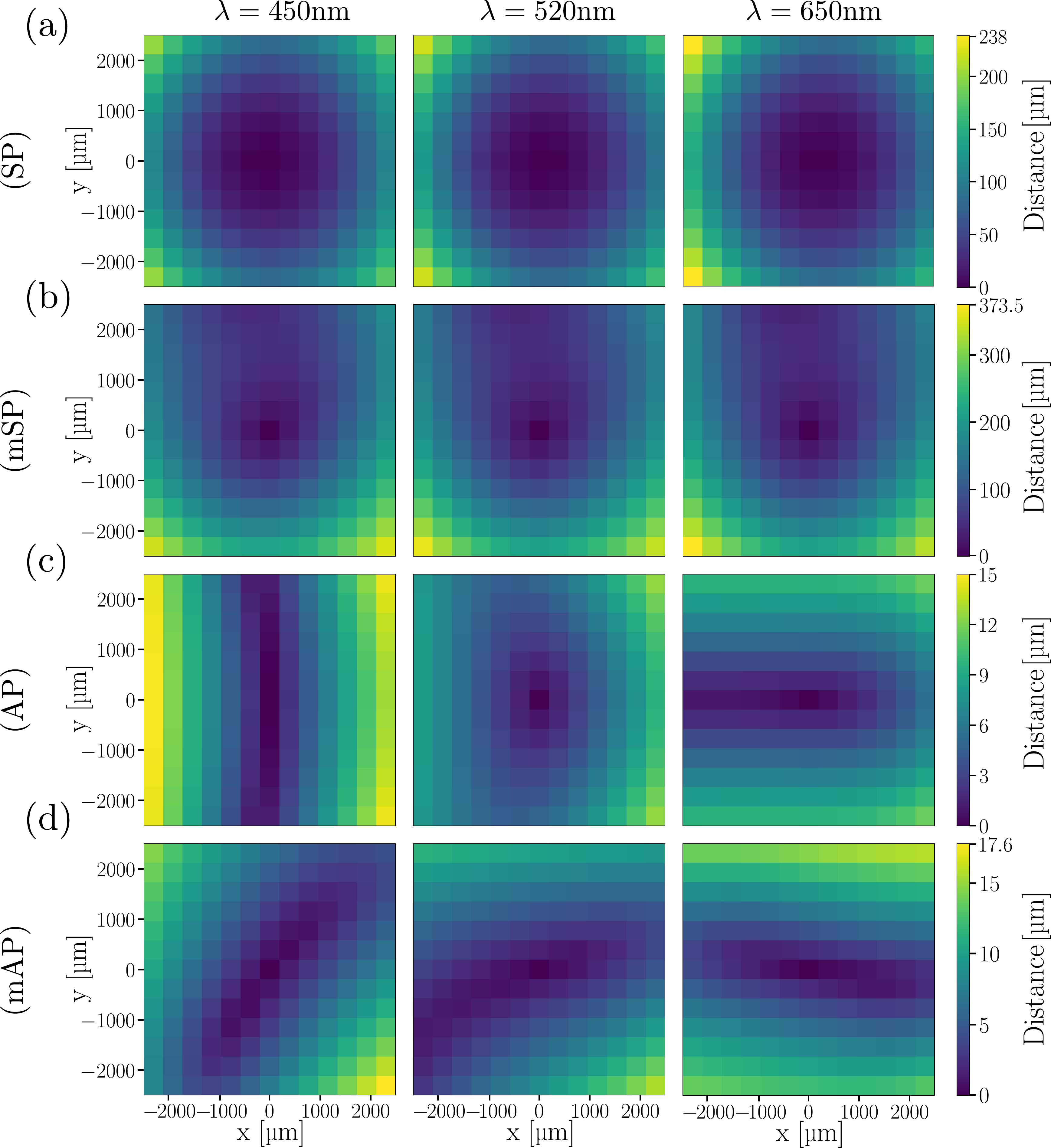}
	\caption{Distortion maps of a regular grid of points traced through the four considered optical systems, for the extrema and center wavelengths (450\,nm, 520\,nm, and 650\,nm). Note the different colorscales between the configurations.}
	\label{fig:distorsions_maps_field}
\end{figure}

In terms of distortions however, both systems are extremely different.
With respect to the (SP) setup, the maximum geometric distortion in (AP) is reduced from $214\,$µm to $6\,$µm and the mean distortion from $75\,$µm to $1.8\,$µm over a $5\times5$\,mm field of view.
Figure\,\ref{fig:distorsions_maps_field} shows the distortion map for both systems (SP) and (AP), and also for the misaligned configurations (mSP) and (mAP).
The interpretation of these distortions in terms of keystone and smile is further discussed in section\,\ref{supp-sec:distortions} and Figure\,\ref{supp-fig:spot_diagram} of the Supplementary.  

This distortion minimization is achieved while maintaining a direct-view geometry with a total deviation $D$ of $0.1$\,mrad. The spatio-spectral spreadings on the detector equal 830\,µm and 813\,µm respectively for the (AP) and (SP) setups. 

Both the commercially available N-BK7 equilateral prism and our custom-designed double-Amici prism assembly exhibit comparable spectral dispersion while presenting different geometric distortions, especially the substantial smile distortion induced by the equilateral prism, ensuring a fair comparison of systems performances in section\,\ref{subsec:results}.

\subsection{Differentiable Simulation for SD-CASSI systems}
\label{subsec:implementation_sd-cassi}

Both single prism and double Amici prism assembly are employed to implement SD-CASSI systems using the differentiable optics design tool dO.

Figure\,\ref{fig:system_layout} shows the two system's configurations in dO.
As stated in section\,\ref{subsec:systems_settings}, (AP) and (mAP) use a single relay lens (see Figure\,\ref{fig:system_layout}-(b)).
Typically, an achromatic and field-corrected relay lens is utilized to ensure high spatial resolution across the field of view and the entire spectral range. 
The  compound relay lens is modeled as an ideal thin lens to isolate the impact of geometric distortions caused by the prism.
This modeling choice avoids conflating prism-induced anamorphic distortions with lens-induced aberrations, ensuring that any observed deformation in the reconstruction arises from the system configuration.
In the single-prism-based configurations (SP) and (mSP), two lenses are on either side of the prism (see section\,\ref{subsec:systems_settings}).
As proposed in~\cite{Zhao:23}, it is advantageous to minimize aberrations early in the setup by employing an achromatic objective lens before the prism and a simple doublet lens afterwards to reduce costs. 
Similar to the approach used for the Amici system, we model the achromatic objective lens as an ideal thin lens and implement a Thorlabs AC254-050-A-ML doublet lens for the second lens, as illustrated in Figure\,\ref{fig:system_layout}-(a).

The misaligned configurations (mSP) and (mAP) have been simulated by rotating the dispersive element by 5° around the $x$-axis as stated in Section\,\ref{sec:design_simulation}.

\subsubsection{Differentiable Ray-Traced Simulations}

dO supports both forward and backward ray-tracing, enabling the optimization of the design of imaging systems and the rendering of scenes through these systems.
It was originally developed for designing free-form optics RGB imaging systems with axial symmetry: adapting it to CASSI systems required modifications due to the loss of axial symmetry caused by the prism.
These modifications included the addition of a prism optical element, based on dO built-in surfaces, and the ability to perform smooth rotations between optical elements, allowing modeling and simulation of the CASSI architecture.

In our modified dO, each optical element is then treated as an independent system, and rays are traced between these systems.
The optical elements are composed of surfaces separated by materials, as in Zemax.
Each element can then be rotated and shifted individually to closely match the design specifications.

The implementation of SD-CASSI designs in dO enables the ray-traced acquisition of hyperspectral and coded hyperspectral images that reflect the given optical system, in contrast to the simplified mathematical models used in~\cite{Wagadarikar:08, Arguello2013, Kittle:10}.
Such models often struggle to account for all distortions and dispersions present in real optical systems.
Our approach is more aligned with the mathematical models of~\cite{Song:22, Zhang2023}.
In addition, it produces spot diagrams through ray tracing and can account for spectral dispersions along the $y$ (vertical) axis and for the spectral dispersions continuity thanks to Monte-Carlo methods.
This is particularly evident in the case of misaligned optical configurations, as seen in the (mAP) and (mSP) configurations (refer to Figure\,\ref{fig:distorsions_maps_field} and Supplementary Figure\,\ref{supp-fig:spot_diagram}) although the Amici-based system is more tolerant to misalignment.

Additionally, dO supports the differentiable optimization of these designs, allowing fine-tuning of rotations, shifts between optical elements, and adjustments of surface distances and configurations.
This functionality allows both the design and operational use of SD-CASSI systems within the same application, facilitating seamless imaging co-design.

\subsubsection{Validation of the Implementation in dO}
\begin{figure}[!t]
	\centering
	\includegraphics[width=\columnwidth]{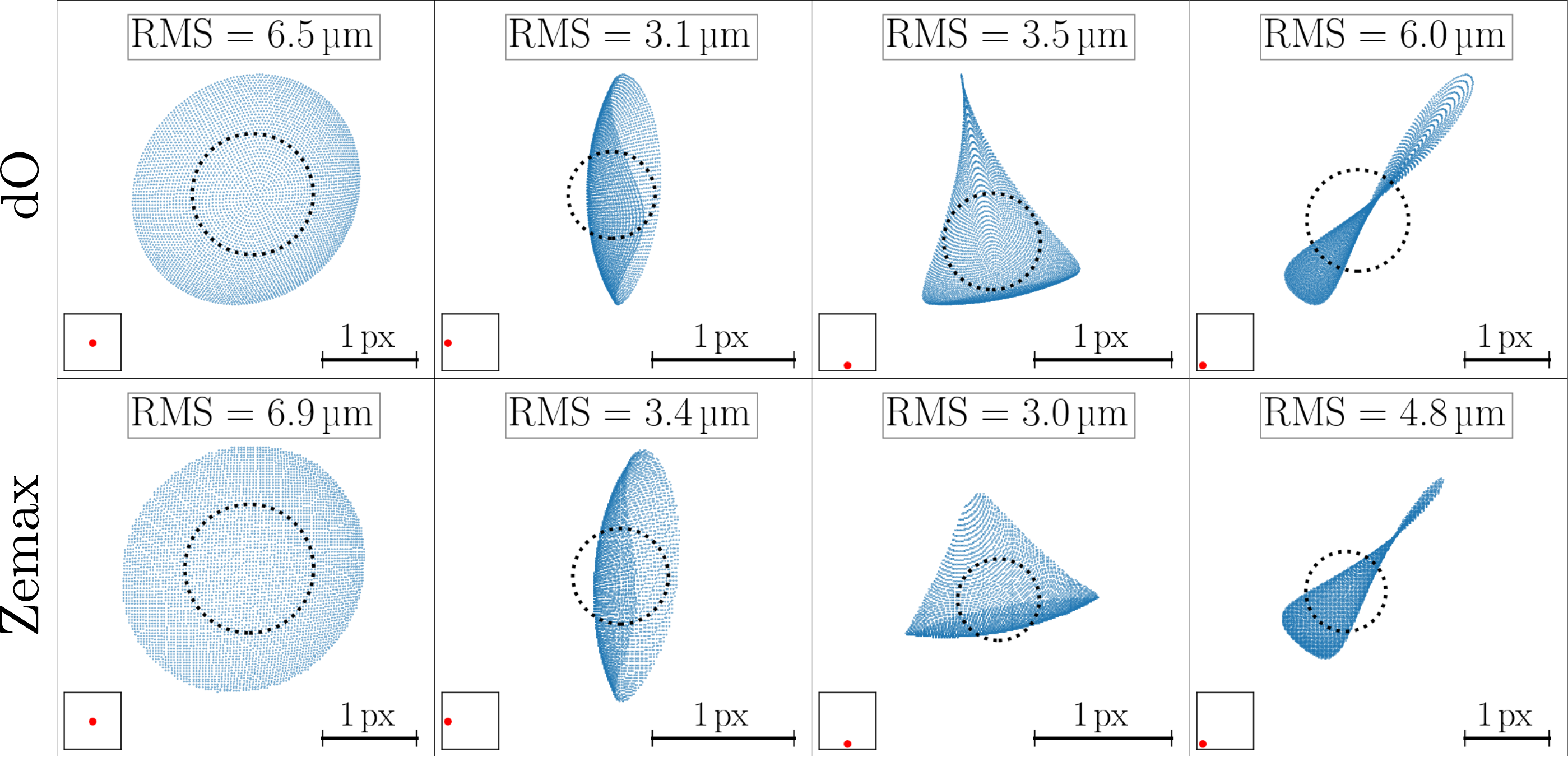}
	\caption{Spot diagrams obtained with the misaligned single prism configuration (mSP) at four positions in the field of view (positions denoted by the red points on the bottom left of each figure). Top: spot diagrams obtained with dO, bottom: spot diagrams obtained with Zemax.  The dotted black circle corresponds to the RMS radius centered on the centroid of the spot diagrams.}
	\label{fig:psf}
\end{figure}

To validate our dO implementation, we compared the spot diagrams from dO and Zemax (considered the groundtruth) of all four configurations across various wavelengths and spatial positions.
Figure\,\ref{fig:psf} shows a selection of four spot diagrams for various positions across the field of view for the (mSP) configuration at a wavelength of $\lambda=650$\,nm (more spot diagrams are presented in the supplementary material, \Cref{supp-fig:psf_amici,supp-fig:psf_single,supp-fig:psf_amici_mis,supp-fig:psf_single_mis}).
Although this configuration exhibits the strongest distortions and is thus the most challenging of the four considered configurations, the spot diagrams modeled with dO are similar to their Zemax counterparts.
The spot diagrams from both implementations (dO and Zemax) are highly consistent, with Root-Mean-Square (RMS) spot size differences below 1.5\,µm, well below the considered 10-µm pixel size, thereby ensuring adequate rendering resolution.

Differences observed in spot diagrams primarily arise from aperture handling which impacts ray distribution: dO assumes a telecentric system without an aperture stop, whereas Zemax includes an aperture object.
Adding an aperture function to dO would require ray-direction calculations for aperture filling, but the benefits would be minimal given the minor discrepancies between both systems.
Additionally, ray sampling methods differ: with dO we employ hexapolar sampling, while Zemax uses grid sampling for spot diagram determination.

\begin{figure}[!t]
	\centering
	\includegraphics[width=0.95\columnwidth]{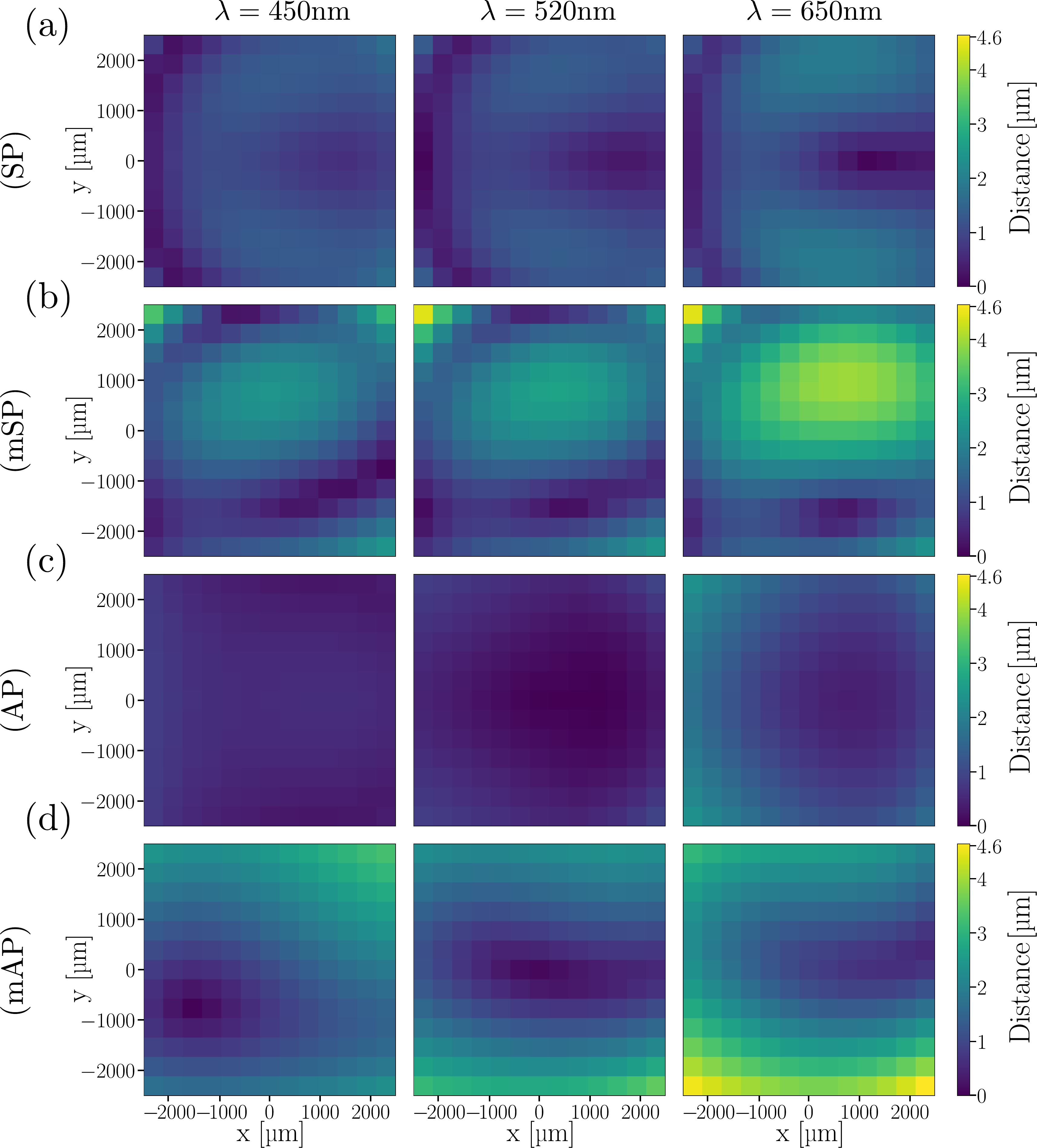}
	\caption{
      Maps of the difference between distortions computed with dO and Zemax, for the extrema and central wavelengths (450\,nm, 520\,nm, and 650\,nm).
      Note the same colorscales across the 4 configurations.
   }
        \label{fig:distorsions_maps_zemax}
\end{figure}

Further validation was conducted by comparing distortions at various wavelengths and positions obtained with both dO and Zemax.
Figure\,\ref{fig:distorsions_maps_zemax} shows the difference between Zemax and dO distortions across the field of view for extrema and central wavelengths.
Results show that distortions are correctly modeled, with mean differences of only 1\,µm across the three wavelengths, a tenth of the actual pixel size.
Minor discrepancies in distortion maps between dO and Zemax also stem from their differing aperture treatments.

\subsection{Coded acquisition rendering}
\label{subsec:image_rendering}

\subsubsection{Forward model}
\label{subsubsec:forward}
To simulate the acquisition of an HSS $\mathbf{H}$ of dimension $H\times W\times K$ --along the vertical, horizontal and spectral directions-- with a given 2D binary mask $\mathbf{M}$ through a SD-CASSI system, we compute the coded scene $\mathbf{H}^c$ as follows:
\begin{equation}
	\label{eq:mask}
\mathbf{H}^c(\cdot, \cdot, k) = \mathbf{H}(\cdot, \cdot, k) \odot \mathbf{M}(\cdot, \cdot), \, \forall k \in \llbracket 0, K \rrbracket
\end{equation}
where $\odot$ is the element-wise product, and $k$ the index along the spectral dimension.
$\mathbf{H}^c$ thus represents information at the entry of our optical system.

The SD-CASSI systems then form the image of a coded scene $\mathbf{H}^c$ through the following forward model.
Each configuration will entail a unique PSF, based on its optics, that will alter the resulting acquisition.
Since our configurations can use both aligned and misaligned prisms, they exhibit spectral dispersions both along $x$ and $y$ dimensions.

In the case of shift-invariant PSFs, the coded acquisition $\mathbf{A}(x, y)$ formed on the detector can be written as:
\begin{equation}
	\mathbf{A}(x, y) = \int\!\mathbf{H}^c(x-s_x(\lambda),y-s_y(\lambda),\lambda) \otimes p(x, y, \lambda)d\lambda
\end{equation}
with $s_x$ and $s_y$ the wavelength dependent dispersion along the $x$ and $y$ axes,
$p(\cdot,\cdot,\lambda)$ the PSF at wavelength $\lambda,$ and $\otimes$ the 2D spatial convolution operator.
Note that when the system is perfectly aligned, $s_y=0$.
However, when accounting for distortions and aberrations, the PSFs are no longer stationary, and vary both with the position on the detector and with the wavelength.
This results in a more realistic model~\cite{Song:22} described in the following equation:
\begin{equation}
	\label{eq:forward_full}
	\begin{split}
	\mathbf{A}(x,y)&= \iiint\! \mathbf{H}^c(x',y',\lambda)\times \\
      &\hspace{-2em}  p(x-x'-s_x(\lambda),y-y'-s_y(\lambda);x,y,\lambda)  dx'dy'd\lambda
	\end{split}
\end{equation}

After spatial and spectral discretization (which requires some approximations), equation\,\eqref{eq:forward_full} writes:
\begin{equation}
	\label{eq:discr_forward_full}
	\begin{split}
	\mathbf{A}(i_d,j_d)&= \sum_k\!\!\sum_{\substack{(i',j')\\\in\,\Omega_{i_d,j_d,k}}}\!\!\!\mathbf{H}^c(i',j',k)\\
	& \hspace{1.0em}\times p(i_d-i'-s_x(k),j_d-j'-s_y(k);i_d,j_d,k)
	\end{split}
\end{equation}

where $\Omega_{i_d,j_d,k}$ represents the pixels of the scene that contribute to a sensor pixel $(i_d,j_d)$  (center position of a detector pixel) at the discrete spectral index $k$.
Such a model is difficult to manage as it requires to know the varying PSFs for all discrete positions $(i_d,j_d)$ and  all wavelengths $k$.
One recent approach\,\cite{Song:22} is to tabulate estimated PSFs with a good-enough spatial and spectral sampling and build a spatially and spectrally varying model in order to approximate the PSFs at any position.
However, this approach requires access to an existing prototype to estimate the PSFs.
Note that works on reconstruction algorithms (e.g.\, \cite{MST, DAUHST, PADUT}) generally use a simplified version of~\eqref{eq:discr_forward_full}, with no PSF (or equivalently a Dirac Delta function as PSF), no dispersion along the $y$ axis ($s_y(k)=0$), and a linear dispersion along the $x$ axis with respect to the wavelength index $k$ (typically $s_x(k) = 2k$ for 28 bands spread over 54 pixels). 
However, the nonlinearity of dispersion is accounted for through a nonlinear relation between the wavelength $\lambda$ and the spectral index $k$ so an irregularly spectrally sampled HSS $\mathbf{H}$ is considered.

\begin{figure}[!t]
	\centering
	\includegraphics[width=0.99\columnwidth]{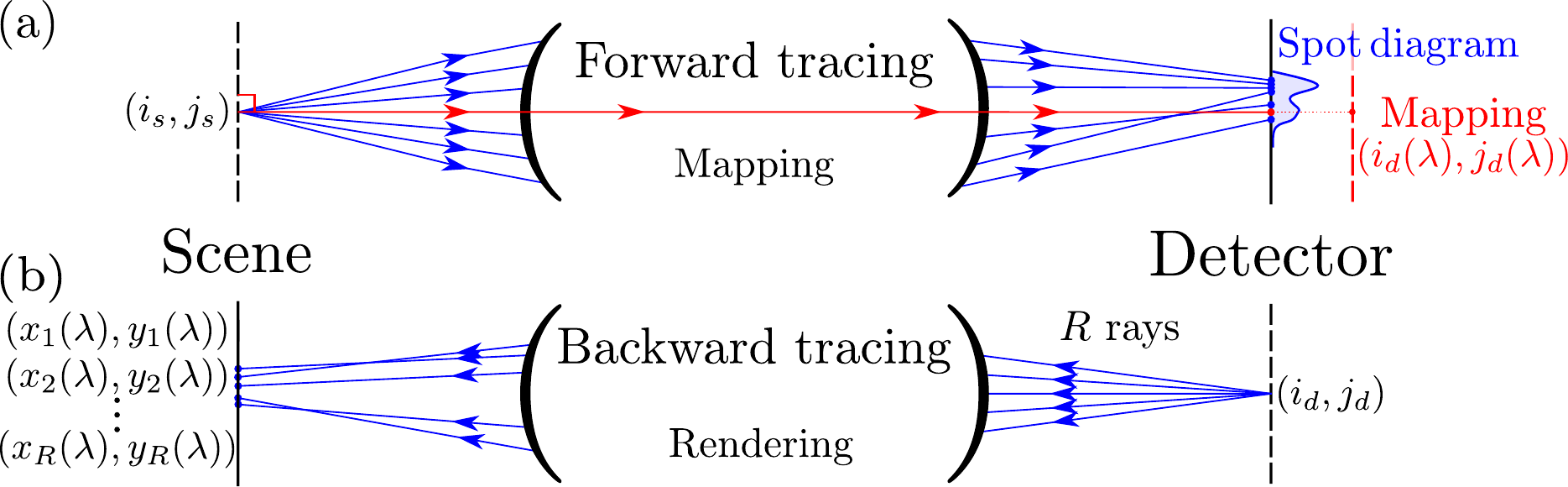}
	\caption{
      Description of the forward and backward ray-tracing (a single wavelength is shown for readability).
		(a) Forward ray-tracing, with a single normal-ray to create the mapping function (red) and with many rays to trace the spot diagrams (blue).
		(b) Backward ray-tracing with $R$ rays, used to render acquisitions by finding contributing voxels in the scene for every detector pixel.
	}
	\label{fig:forward_backward}
\end{figure}

Contrarily, here we focus on CASSI systems at the design stage, where a prototype is yet to be built.
Therefore, we propose to use a ray-tracing approach similar to our rendering process.
Instead of weighting the coded scene $\mathbf{H}^c$ at pixel location $(i',j')$ with an approximated PSF, we propose to launch $R$ rays backward from each pixel of the detector to the coded scene for each wavelength.
The forward model corresponding to such a ray-tracing approach writes:
\begin{equation}
	\label{eq:raytraced_forward}
	\mathbf{A}(i_d,j_d) = \sum_{k} \frac{1}{|\Omega_{i_d,j_d,k}|} \sum_{r=1}^{R} \mathbf{H}^c(x_{r}(k), y_{r}(k), k)
\end{equation}
where $(x_{r}(k), y_{r}(k))$ corresponds to the pixel position of a ray $r$ on the scene plane after being traced backward from the detector pixel $(i_d,j_d)$ at wavelength index $k$ (wavelength $\lambda$, see Figure\,\ref{fig:forward_backward}-(b)) and $|\Omega_{i_d,j_d,k}|$ is the number of such rays that actually reach the scene plane -- some rays might be blocked somewhere along the optical path.
$\mathbf{H}^c(x_r(k), y_r(k), k)$ is approximated by interpolation from the values of neighboring pixels in the scene at wavelength $k$.
In practice, we have chosen to use linear interpolation, but other interpolation schemes could be used, at the cost of a higher computational overhead.
This equation is close to Eq.\,\eqref{eq:discr_forward_full} since we estimate which scene voxels contribute to the rendered image on a given detector pixel $(i_d, j_d)$, but the weighting is deduced from the backpropagated rays instead of the PSF.
Note that with such a ray-tracing approach, the nonlinear dispersion of the prisms and the distortions are inherently accounted for in the model as the rays are deviated by the optical elements of the system. 
Therefore our model does not require an HSS input that is irregularly sampled in the spectral dimension, but rather a regularly sampled one.
It is worth noting that using backward ray-tracing to describe the forward propagation is a well established approach for rendering, although rarely used for HS scenes.
This process enables the efficient computation of the acquisition $\mathbf{A}$ directly from the simulator, without requiring a pre-existing prototype.

\subsubsection{Rendering process}
\label{subsubsec:rendering}
Our model of the four optical configurations enables HSS rendering through backward ray-tracing.
Each spectral plane is rendered individually, and acquisition is performed by summing along the spectral dimension as described in equation~\eqref{eq:raytraced_forward}.
We render scenes from the hyperspectral datasets CAVE~\cite{CAVE} and KAIST~\cite{Choi:17}, preprocessed as in~\cite{MST} considering their 28 spectral bands to be evenly distributed from 450\,nm to 650\,nm.
Due to the systems' nonlinear spectral spread of 830\,µm on the detector and the spatial sampling of 10\,µm, the input spectra are spread over a range of 83 pixels of the detector.

To obtain a correct rendering from these 28-bands data, we had to spectrally upsample the data, otherwise some pixels would not receive any signal.
This necessity is highlighted in Supplementary Figure\,\ref{supp-fig:diff_sampling}, where we show the rendered acquisitions of a scene upsampled with different upsampling factors.
The smallest integer upsampling factor $n$ such that $28\times n\geq 83$ is $n=3$: we upsample with $n=4$ to alleviate missing signal errors that could occur due to Monte-Carlo sampling with a small amount of rays.
The upsampling is done through linear interpolation.
We therefore use these upsampled scenes in equation\,\eqref{eq:mask}.
Note that this spectral upsampling would not be needed for data with a higher spectral resolution, and has only been performed because we use datasets with a small number of spectral bands.
The shortcoming of this upsampling is that the spectra may be smoother than they actually are.

Rendered HSSs therefore contain $4 \times 28 = 112$ spectral bands.
For a fast rendering of the $512 \times 512 \times 112$ hyperspectral scenes, we limit ray tracing to $R=20$ rays per pixel per wavelength (see equation\,\eqref{eq:discr_forward_full}).
Following the modifications done to dO, rendering is performed sequentially, progressing backwards through each optical element and independently for each wavelength.
To further mitigate quantization errors from Monte-Carlo sampling arising from tracing a small amount of rays, the rendered planes are convolved with a smoothing kernel.
We convolve each spectral plane with an Airy kernel, varying according to $\lambda$, whose diameter equals 2.5 pixels ($= 25$\,µm) at 520\,nm, adequately smoothing the rendering through convolution.

\subsubsection{Validation of the Rendering Method}
We validate our rendering method with a vertical, 1-pixel-wide, single-slit mask, for which the SD-CASSI systems emulate an imaging prism spectrometer.
As both aligned configurations (AP) and (SP) disperse the spectrum of the imaged slit horizontally along the $x$ axis, the spectrum of each point along the slit can be directly read on the corresponding horizontal line on the detector.
Since distortions mainly appear in the (SP) configuration, we will only consider this harder case.

\begin{figure}[!t]
	\centering
	\includegraphics[width=0.95\columnwidth]{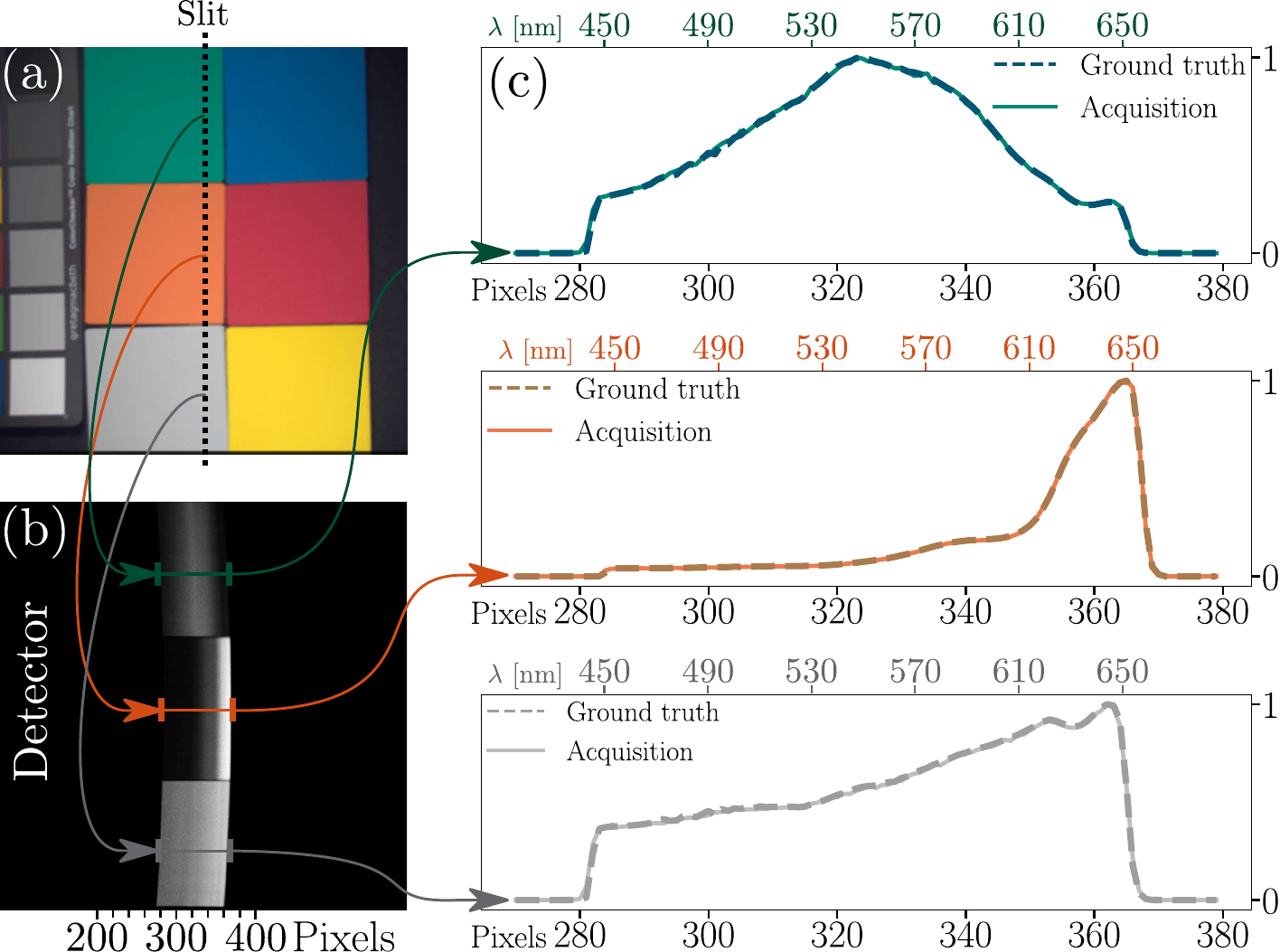}
	\caption{
       	 Spectra extracted from a rendered single-slit acquisition with the (SP) configuration compared to ground truth spectra retrieved from the HSS.
         The spectrum of each pixel along the slit is horizontally spread along the rows of the detector.
		 (a) RGB image of the scene and location of the slit (dashed line).
		 (b) Rendered acquisition with the (SP) configuration.
		 (c) Acquired spectra (solid lines) in the 3 regions with constant color (green, orange and gray) compared to their ground truth counterparts (dashed lines).
         For each curve, the bottom scale is the position on the detector (in pixel), the top one is the corresponding wavelength.
    }
	\label{fig:spectra}
\end{figure}

The test scene selected from the CAVE dataset is shown in Figure\,\ref{fig:spectra}-(a), together with the chosen slit position (dashed line).
The slit is located at the center of the field of view, crossing 3 regions (green, orange and gray squares) with a constant spectrum.
The acquisition shown in Figure\,\ref{fig:spectra}-(b) clearly illustrate the spatial dispersion and the effects of distortion.
Our aim is to have a fair comparison of the spectra extracted from the acquisition and the ground truth, retrieved  from the HSS, as 
illustrated in Figure\,\ref{fig:spectra}-(c).
On one hand, to account both for the $y$-axis spread and the rendering with a small amount of rays, we average 40 rows of the acquisition in each region.
On the other hand, the ground truth spectra are derived from the HSS, accounting for the dispersion model and the convolution with the Airy kernel used in the rendering process.
Moreover, we upsample the ground truth spectra to 280 wavelengths to account for their continuity and to reduce the quantization errors that would otherwise appear in the comparison, with no physical meaning.

Results presented in Figure\,\ref{fig:spectra}-(c) show a close match between the acquisitions (solid lines) and the ground truth spectra (dashed lines).
Thus, our model effectively simulates a SD-CASSI system, enabling ray-traced acquisition simulations that align with those from a physical system.

We created a fork of dO including all our modifications that can be accessed at \url{https://github.com/lpaillet-laas/DiffOptics}.

Additionally, another repository accessible at \url{https://github.com/lpaillet-laas/DiffCassiSim}, contains this fork of dO together with all the code for the processing done in Section\,\ref{sec:performance_evaluation}.

\section{Hyperspectral cube reconstructions}
\label{sec:performance_evaluation}

\subsection{Overview}
We demonstrate here that
for the four considered configurations, a comparable amount of information is encoded in the acquisitions and that it can be retrieved by state-of-the-art approaches when the system model is known and considered, hence exhibiting the marginal impact of distortions and misalignments.

For this purpose, the hyperspectral cube reconstruction from coded acquisition is an ideal test case.
It consists in reconstructing a scene $\mathbf{H}$ from a coded acquisition $\mathbf{A}$.
Deep learning methods~\cite{Wang2018, Choi:17, Wang:19} being now the best solution for HSC reconstruction, we apply five recent state-of-the-art algorithms, and assess the reconstruction quality with a series of usual metrics.
The considered networks are Deep Gaussian Scale Mixture Prior (DGSMP)~\cite{DGSMP}, Mask-guided Spectral-wise Transformer (MST)~\cite{MST}, Degradation-Aware Unfolding Half-Shuffle Transformer (DAUHST)~\cite{DAUHST}, Residual Degradation Learning Unfolding Framework (RDLUF)~\cite{RDLUF}, and Pixel Adaptive Deep Unfolding Transformer (PADUT)~\cite{PADUT}.

Our evaluation uses the two public HSS datasets CAVE~\cite{CAVE} and KAIST~\cite{Choi:17}.
The CAVE dataset includes 32 HSSs of size $512 \times 512 \times 31$, while the KAIST dataset includes 30 HSSs of size $2704 \times 3376 \times 31$.
We considered the same datasets than the used algorithms, spectrally interpolated following~\cite{TSANet} in order to be consistent with their model, but considered the spectral bands to be evenly distributed over the range 450$\,$nm to 650$\,$nm. 

To fit the field of view ($\simeq 5 \times 5$\,mm$^2$) with a pixel size of 10\,µm, each HSS in the datasets was cropped to reach a final size of $512 \times 512 \times 28$.
It should be noted that oversampling up to 112 spectral bands of the HSS was necessary for the rendering step, as explained in\,\ref{subsubsec:rendering}, but the HSCs reconstructed by the algorithms have 28 bands, like the original dataset.	
Following the setups of the referenced networks, models were trained on the CAVE dataset and tested on 10 scenes taken from the KAIST dataset.

\subsection{Spatio-Spectral Mapping for Reconstruction Algorithms}
\label{sec:mapping}
Most networks leverage \emph{a priori} information from the acquisition to reconstruct the hyperspectral cube.
The considered reconstruction algorithms contain a simplified propagation model to achieve better results than if they were purely data-driven.
More precisely,  the used networks all consider as input the binary mask $\mathbf{M},$ an initial HSC $\mathbf{I},$ and leverage \emph{a priori} information from the acquisition model through a mapping function $f$:
\begin{equation}
	f(i_s, j_s, k) = (i_d, j_d)
	\label{eq:mapping}
\end{equation}
which gives the position on the detector $(i_d, j_d)$ impacted by the intensity at a given spectral index $k$ of the input scene at position $(i_s, j_s).$
This mapping function is used in different steps of the networks, but also for the initialisation as
\begin{equation}
	\mathbf{I} = \mathbf{A} \circ f.
	\label{eq:mapinginiti}
\end{equation}

For example, for perfectly aligned systems without distortions, a scene $\mathbf{H}$ with dimensions $H \times W \times K$ is rendered onto an acquisition $\mathbf{A}$ of size $H \times (W + S)$, where $S$ denotes the spectral spread (in pixels) of the optical system.
Thus, for the simplified model with $s(k)=2k,$ the mapping function is: 
\begin{equation}
	f(i_s,j_s,k)=(i_s,j_s+2k).	
	\label{eq:mappingsimple}
\end{equation}
and the initialization Equation\,\eqref{eq:mapinginiti} amounts to:
\begin{equation}
	\mathbf{I}(:, :, k) = \mathbf{A}(:, s(k):W + s(k)), \, \forall k \in \llbracket 0, K \rrbracket.
\end{equation}
resulting in an initialized cube of dimensions $H \times W \times K$.

With negligible distortions, all information from the scene at wavelength index $k$ is indeed captured in a rectangular area of size $H \times W$ starting at the position $i_d=s(k)$.
However, as shown in Figure\,\ref{fig:distorsions_maps_field} and in Figure\,\ref{supp-fig:spot_diagram} of the supplementary material, some of our configurations exhibit strong spatial distortions, so information from each wavelength is not represented in a rectangular area.
In the misaligned configurations (mAP) and (mSP), the acquisition $\mathbf{A}$ also displays dispersion along the $y$ axis, resulting in an acquisition of dimensions $(H + S_x) \times (W + S_y)$, where $S_x$ denotes the spectral spread (in pixels) along the $x$ axis, and $S_y$ the spectral spread along the $y$ axis.
Therefore, for such configuration, the mapping function is not as simple as Equation\,\eqref{eq:mappingsimple}.
To build the corresponding mapping function, we forwardly ray trace a single ray parallel to the optical axis -- a normal-ray -- for each position in the scene and each wavelength (red ray in Figure\,\ref{fig:forward_backward}-(a)).

\begin{figure}[!t]
	\centering
	\includegraphics[width=0.75\columnwidth]{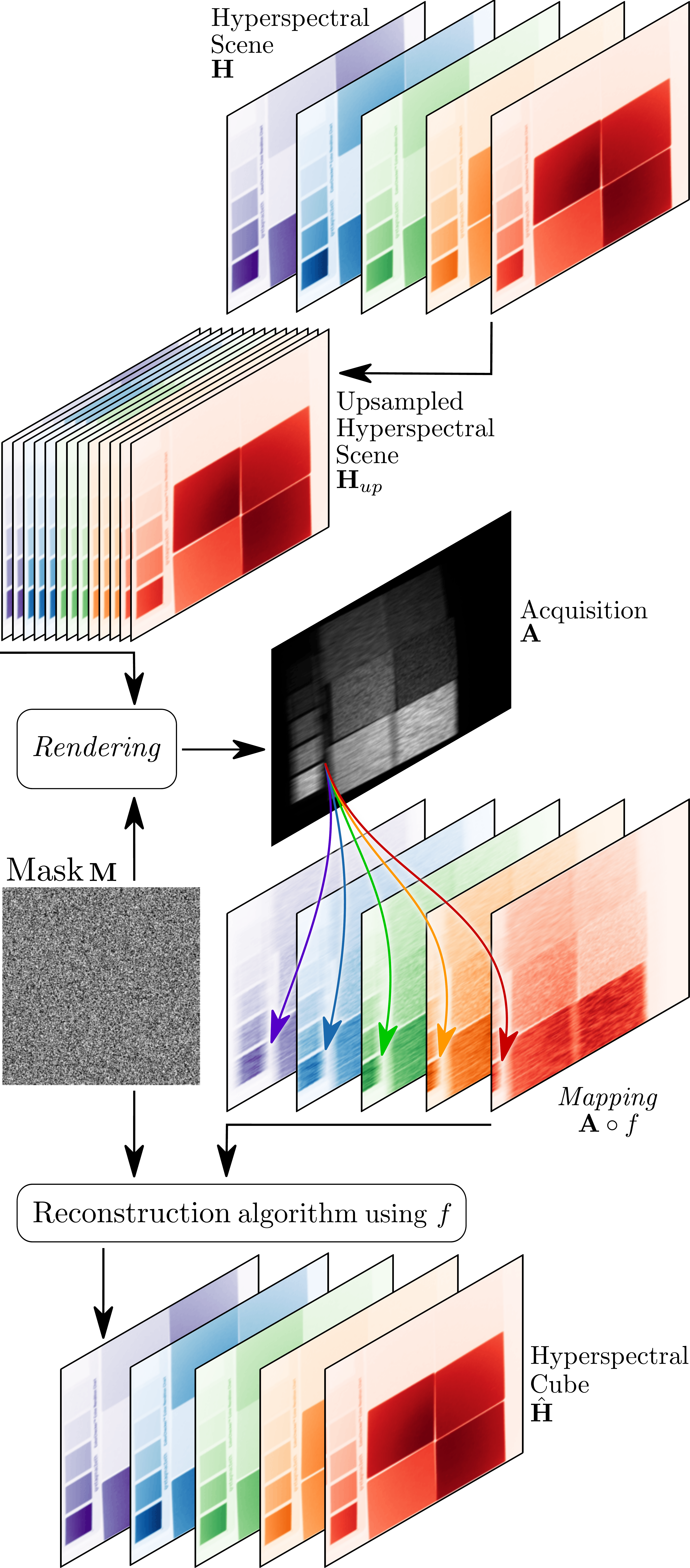}
	\caption{
      Workflow to reconstruct HSCs with a given optical system and a given reconstruction algorithm.
      The scene $\mathbf{H}$ is upsampled into $\mathbf{H}_{up}$ and then acquired with the optical system and the mask $\mathbf{M}$ thanks to our rendering with dO.
      The acquisition is then mapped to an initialization using the model through $f$.
      This serves as an input to a reconstruction algorithm yielding $\hat{\mathbf{H}}$.
      The mapping function is also used in the algorithm.}
	\label{fig:overview}
\end{figure}

As illustrated in Figure~\ref{fig:overview}, to ensure accurate computation, we have modified the algorithms to incorporate the mapping function $f.$ 
For each configuration, function $f$ is adapted to the model of the optical system in use,  encompassing known distortions, and to the 28 spectral bands of the datasets.

\subsection{Networks Implementation}
The networks were adapted to accommodate our rendered acquisitions.
Originally designed to reconstruct $256 \times 256 \times 28$ HSCs, the networks were modified to reconstruct $512 \times 512 \times 28$ HSCs required by our configurations.
Additionally, most networks were previously hardcoded for simplified optical setups assuming linear spectral dispersion with a slope of 2 pixels (20\,µm in our case) across 560\,µm.
However, our configurations exhibit unique nonlinear spectral spread across 830\,µm.
The networks were therefore adapted with the mapping function $f$ to process the acquisitions, as illustrated in Figure\,\ref{fig:overview}.
All reconstructions were performed on noise-free simulated acquisitions, the only noise comes from tracing only 20 rays per pixel per wavelength.

For all HSSs, a unique binary mask with a random opening ratio of 0.5 was used.
Consistent with original methods, we use the Adam optimizer and both learning rates and scheduler schemes were maintained.
Each network and configuration was trained for 500 epochs, utilizing random cropping, flipping, and rotation for data augmentation.
The batch size was set to 1, with gradient accumulation every 4 batches, and training conducted on NVIDIA RTX A100 and A6000 GPUs.

Training loss was calculated as the Root Mean Square Error (RMSE) between the ground truth HSSs and reconstructed HSCs.
Evaluation metrics include RMSE, Peak Signal-to-Noise Ratio (PSNR), Structural Similarity Index Measure (SSIM)~\cite{SSIM}, and Spectral Angle Mapper (SAM)~\cite{SAM}. The SAM metric allows to evaluate the spectral accuracy of the reconstructions.
The definition of these metrics can be found in section\,\ref{supp-sec:recons} of the Supplementary material.

\subsection{Results}
\label{subsec:results}

\begin{table}[h!]
	\resizebox{\columnwidth}{!}{
	\begin{small}
		\centering
		\begin{tabular}{>{\bfseries}c l c c c c}
			\toprule
			\rowcolor{gray!20}
			& & (AP) & (SP) & (mAP) & (mSP)\\ 
			\midrule
			\multirow{5}{*}{\begin{tabular}{c}RMSE $\downarrow$\\
					($\times 10^{-3}$)\end{tabular}} & DGSMP & \textbf{31.3} & 32.8 & 31.8 & 34.4 \\
			& MST & 27.3 & \textbf{26.5} & \textbf{26.5} & 27.0 \\
			& DAUHST & 20.7 & 19.6 & \textbf{19.5} & 20.3 \\
			& RDLUF & \textbf{22.6} & 24.1 & 23.8 & 23.7 \\
			& PADUT & 21.4 & \textbf{19.6} & 21.6 & 21.8 \\
			\midrule
			\multirow{5}{*}{PSNR $\uparrow$} & DGSMP & \textbf{30.6} & 30.3 & 30.5 & 29.9 \\
			& MST & 31.9 & \textbf{32.2} & 32.1 & 32.0 \\
			& DAUHST & 34.4 & 34.7 & \textbf{34.9} & 34.4 \\
			& RDLUF & \textbf{33.5} & 32.9 & 33.0 & 33.0 \\
			& PADUT & 34.1 & \textbf{34.7} & 34.0 & 33.8 \\
			\midrule
			\multirow{5}{*}{\begin{tabular}{c}SSIM $\uparrow$\\
					$[0-1]$\end{tabular}} & DGSMP & \textbf{0.892} & 0.883 & 0.886 & 0.879 \\
			& MST & 0.910 & \textbf{0.914} & 0.912 & 0.911 \\
			& DAUHST & 0.942 & \textbf{0.945} & \textbf{0.945} & 0.939 \\
			& RDLUF & \textbf{0.930} & 0.922 & 0.926 & 0.926 \\
			& PADUT & 0.937 & \textbf{0.943} & 0.936 & 0.935 \\
			\midrule
			\multirow{5}{*}{\begin{tabular}{c}SAM $\downarrow$\\
					$[0-1]$\end{tabular}} & DGSMP & \textbf{0.058} & 0.062 & 0.059 & 0.065 \\
			& MST & 0.055 & \textbf{0.052} & 0.055 & 0.055 \\
			& DAUHST & 0.048 & 0.048 & \textbf{0.047} & 0.050 \\
			& RDLUF & \textbf{0.053} & 0.055 & 0.054 & \textbf{0.053} \\
			& PADUT & \textbf{0.048} & \textbf{0.048} & 0.049 & \textbf{0.048} \\
			\bottomrule
		\end{tabular}
	\end{small}
	}
	\caption{Average RMSE, PSNR (in dB), SSIM and SAM on all test scenes, for each optical system and reconstruction algorithm. Bold numbers correspond to the best configuration for a given metric and algorithm.}
	
	\label{tab:results_metrics}
\end{table}

Following the setups of the referenced networks, models were trained with our four configurations on the CAVE dataset and tested on 10 scenes taken from the KAIST dataset.
The quantitative results are presented in Table\,\ref{tab:results_metrics}.
SAM has been normalized between 0 and 1 to be in the same range as SSIM.
We compare the evaluation metrics across all four configurations for each state-of-the-art reconstruction algorithm.
The evaluation metrics are slightly worse than those reported in corresponding papers.
This occurs because we adapted the networks to process $512 \times 512 \times 28$ HSCs, although they were originally designed to reconstruct $256 \times 256 \times 28$ HSCs.
Moreover, our spectral resolution leads to a spreading range of 83 pixels against the 54 usually considered for the simplified SD-CASSI models.
Finally, we only passed through the preprocessed dataset~\cite{MST} once per epoch.
Thus, achieving the same performance as the reference papers might require either more training epochs or a deeper architecture to better capture small spatial variations and bigger spectral spreading.
Some modifications to the learning procedures, or even to the algorithms themselves, could have improved the results for each algorithm. However, as our aim is not to compare algorithms with each other but, for a given algorithm, to compare configurations with each other, such optimization is not crucial.

As seen in Table\,\ref{tab:results_metrics}, for a given reconstruction algorithm, all four evaluation metrics show no significant difference across all four configurations.
For all reconstruction algorithms, the SSIM difference does not exceed 0.013, and the PSNR difference does not exceed 0.9\,dB.
For further details, the metrics for each scene can be seen in the supplementary materials (see Supplementary~\Cref{supp-tab:results_dgsmp,supp-tab:results_mst,supp-tab:results_dauhst,supp-tab:results_duf,supp-tab:results_padut}).
These non-significant differences indicate that the encoded information quality remains consistent across all four configurations, despite distortions and misalignments in some configurations.
Moreover, the configuration that yields the best results varies depending on both the metric and the algorithm used, with each configuration performing best at least once. This means there is no single overall best configuration.
This further validates that processing results mainly depend on the quality of the processing algorithm, rather than the optical system, provided the system is accurately modeled and its distortions are known.
Thus, we can relax certain constraints on optical design while achieving the same processing performance.

\subsection{Ablation Study}
\label{subsec:ablation}
An ablation study was conducted to evaluate the impact of rendering accuracy and mapping on HSC reconstruction.
The study was performed on two distortion-inducing configurations: (SP) and (mSP), and exclusively employed the PADUT algorithm, as it provided the highest reconstruction quality.

In the first experiment, we assessed the effect of incorrect rendering during the training phase.
PADUT was trained using a simplified rendering, and reconstruction quality metrics were evaluated with our rendering.
The simplified model renders spectral planes using the (AP) system, which introduces negligible distortions and serves as the baseline for standard simplified frameworks~\cite{Wagadarikar:09, Arguello2013, Kittle:10}. 
To avoid bias in reconstruction, the spatio-spectral spreading $S$ curve was set to match that of the propagation model, minimizing minor differences between the systems (see Figure\,\ref{fig:dispersion_curves}).
Spatial shifts were thus applied to the rendered spectral planes to account for spatio-spectral spreading mismatches before summing them to form the acquisitions.

In the second experiment, the mapping function $f$ was entirely removed from the workflow, meaning a simplified propagation model is used both during the training and the test.
Indeed, only the horizontal spatio-spectral spreading characteristics of the configurations were used for initial reconstruction and during the algorithm's processing, to account for the correct dispersion and to prevent bias.
This experiment isolates the influence of the mapping function on reconstruction quality.
Both (SP) and (mSP) configurations were considered to account for distortions and misalignments, with the mapping excluded during training and testing.

\begin{table}
	\centering
	\begin{tabular}{c c c c c}
		\toprule
		\rowcolor{gray!20}
		System & Training & Test & PSNR $\uparrow$ & SSIM $\uparrow$ \\
		\midrule
		\multirow{3}{*}{(SP)} & Simple & Simple &  33.6 & 0.932 \\
		& Simple & Ours & 22.2 & 0.730 \\
		& Ours & Ours & \textbf{34.7} & \textbf{0.943}\\
		\midrule
		\multirow{3}{*}{(mSP)} & Simple & Simple &  \textbf{34.3} & \textbf{0.937} \\
		& Simple & Ours & 17.7 & 0.568 \\
		& Ours & Ours & {33.8} & {0.935}\\
		\bottomrule
	\end{tabular}
	\caption{Rendering ablation result, with PADUT algorithm.}
	\label{tab:ablation_results_rendering}
\end{table}
\begin{table}
	\centering
	\begin{tabular}{c c c c}
		\toprule
		\rowcolor{gray!20}
		System & Mapping & PSNR $\uparrow$ & SSIM $\uparrow$ \\
		\midrule
		\multirow{2}{*}{(SP)}& Without  & 33.6 & 0.936 \\
		& With & \textbf{34.7} & \textbf{0.943} \\
		\midrule
		\multirow{2}{*}{(mSP)} & Without & 32.2 & 0.916 \\
		& With & \textbf{33.8} & \textbf{0.935} \\
		\bottomrule
	\end{tabular}
	\caption{Mapping ablation result, with PADUT algorithm.}
	\label{tab:ablation_results_mapping}
\end{table}

Results for the first experiment are presented in Table\,\ref{tab:ablation_results_rendering}.
They demonstrate that using different rendering models between the training and test phase significantly degrades reconstruction quality.
However, using the same rendering method for both phases gives similar performances for all configurations, as seen in section\,\ref{subsec:results}.
This highlights the fact that reconstruction algorithms cannot be effectively trained on acquisitions rendered using simplified models with the expectation of robust performance when tested on systems employing more realistic, non-simplified rendering processes, or on real acquisitions from CASSI prototypes.

In the second experiment (see Table\,\ref{tab:ablation_results_mapping}), the removal of the mapping function $f$ from the reconstruction process only caused a slight deterioration in reconstruction quality.
We hypothesize that the limited impact is due to the narrow spatio-spectral spread difference between consecutive wavelengths, which allows missing spatial information to be interpolated from neighboring wavelengths during training.
Still, excluding a correct mapping during the training phase notably reduces the achievable quality metrics, thereby amplifying the performance gap relative to systems without distortions or misalignments.

Ultimately, rendering accuracy during both training and test has a much larger influence on reconstruction quality than the inclusion of a correct mapping function.
However, a proper mapping still contributes to improved algorithm performance, providing a non-negligible enhancement in reconstruction accuracy.
Both steps are therefore important to reach the best reconstruction quality with a given algorithm.

\subsection{Tolerance to misalignment}

\begin{figure}[!h]
	\centering
	\includegraphics[width=0.99\columnwidth]{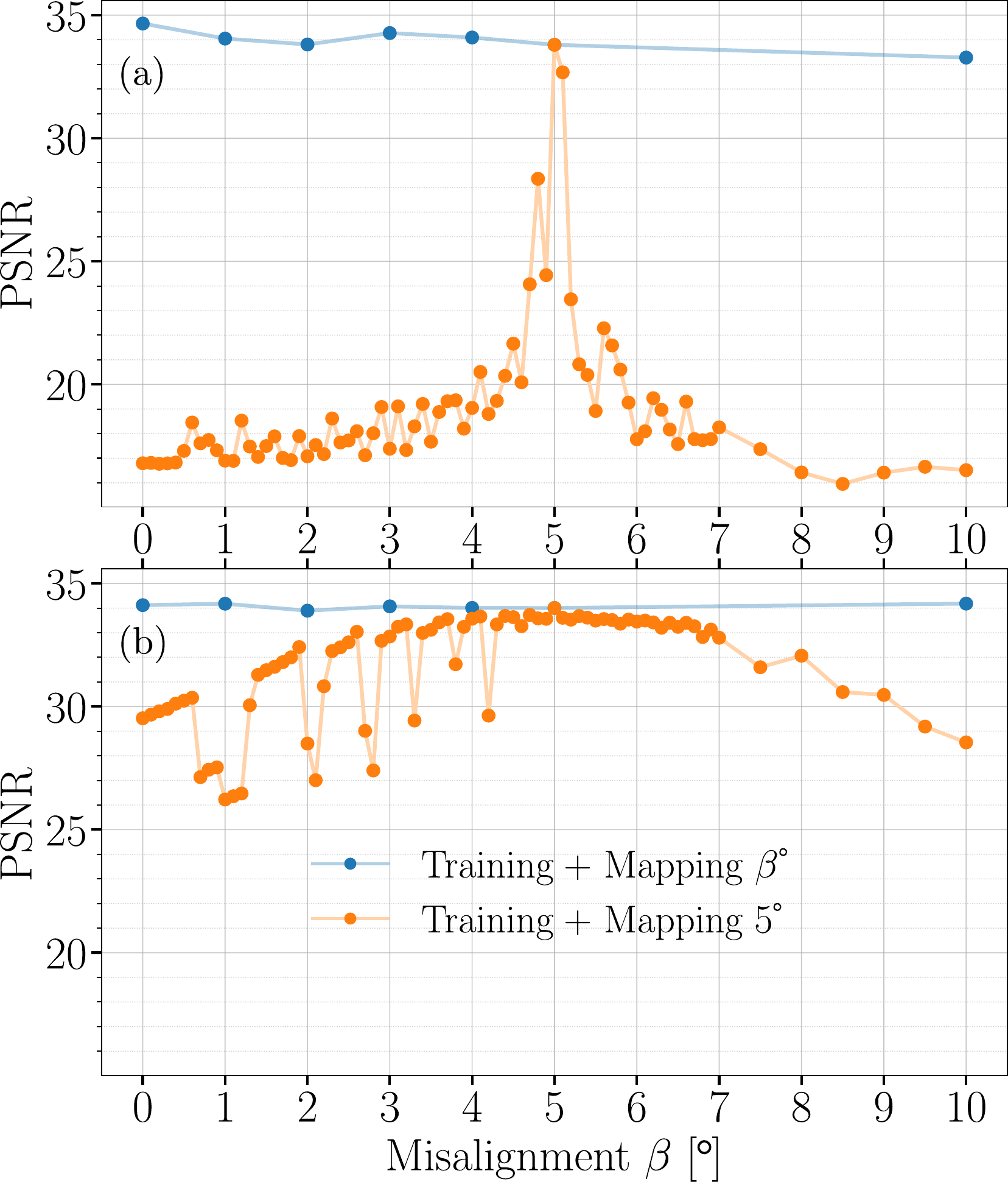}
	\caption{
      Impact of testing misalignment angle $\beta$ (a) for (mSP), (b) for (mAP), using PADUT.
      In blue, the correct misalignment is used for training and mapping;
      in orange, for both training and mapping function a fixed misalignment of 5° is used.
		}
	\label{fig:psnr_misalignment}
\end{figure}

We further investigated the impact of misalignment on system performance by studying the evolution of reconstruction quality with increasing misalignment angle $\beta$.
We compared the impact of a known versus an unknown misalignment.
To this end, we trained PADUT algorithm using (mSP) and (mAP) configurations with a misalignment $\beta$ increasing from 0° to 10°.
We then compared the quality of the reconstruction (using PSNR metric) for two testing scenarios reported in Figure\,\ref{fig:psnr_misalignment}:
a known misalignment scenario using both training and mapping function corresponding to the actual misalignment $\beta$ (blue curve), 
and an unknown misalignment scenario using both training and mapping function at a fixed value of 5°, but testing with a misalignment $\beta$ (orange curve).

Under the first scenario (blue curve), when the misalignment is known and thus both the training and mapping function match the correct misalignment $\beta$, the reconstruction is highly tolerant to misalignment for both (mSP) and (mAP) configurations.
Indeed, over the whole misalignment range, (mSP) exhibits a PSNR drop lower than 1.4\,dB -- thus barely significant -- while (mAP) shows an almost flat PSNR curve.
This shows that when using a correct modeling of the image formation during both training and testing, high and consistent quality reconstruction can be achieved over a wide range of misalignments.

However, (mSP) and (mAP) have different behaviors for the second scenario (orange curve) where the misalignment is unknown and thus we consider a fixed reference misalignment of $\beta=5$° for training and mapping.
While they both show significantly lower PSNR when moving away from the reference misalignment, the tolerance is much narrower and the performance drop higher for (mSP) than (mAP).
Indeed, for a change of misalignment of around $\pm0.5$° around 5°, the PSNR of (mSP) decreases by $\simeq12\,$dB while that of (mAP) stays almost unchanged (the biggest change in that range being lower than $1\,$dB).

These findings indicate that for both configurations, using a mapping function and appropriate training can largely improve the tolerance to misalignment
and yield high reconstruction quality over a wide range of misalignments, even for simple configurations like (mSP) provided that the misalignment is known.
But if the misalignment is unknown, the performance results can decrease depending on the configuration.

\section{Conclusion}

We focused on the design and performance evaluation of Coded Aperture Snapshot Spectral Imaging (CASSI) systems. Our work aimed to bridge the gap between optical hardware and computational processing by providing ray-traced simulations and analyses that consider the complexities of actual optical setups.

Our main contribution was to show that even with simple optical systems, significant distortions and large misalignments arising from the prism, modern algorithms can reconstruct high quality hyperspectral scenes, provided that the image formation model used for training and reconstruction is realistic enough and distortions are known.

In order to reach that goal, our first contribution was the implementation of coded aperture hyperspectral optical systems within a differentiable ray-tracing framework. This enables to render synthetic coded hyperspectral images incorporating optical distortions and aberrations.
By leveraging this framework, we could simulate optical designs more precisely than with simplified mathematical models, despite using ideal thin lenses and not completely simulating PSFs because our model does not include wave optics.
Secondly, a propagation model was utilized to map 2D coded measurements to estimations of hyperspectral cubes, serving both as the initialization and throughout the reconstruction process.
This approach allowed for improved reconstruction by accounting for the exact optical characteristics of the configuration, including geometric distortions and spectral dispersion.
This workflow can be generalized to a great variety of systems with our framework, given the propagation model is known.
Moreover, when working with a prototype, one can also derive the mapping function from acquired experimental PSFs.
Thirdly, we demonstrated that geometric distortions and misalignments in CASSI systems have a marginal impact on reconstruction performance when they are known and accounted for.
We conclude that the same amount of information is transmitted regardless of the system used.
However, unknown distortions and misalignment are still a limitation to our model.
	
Looking forward, this work yields the comparison of different CASSI system performances using information theory measures, facilitating more informed design choices~\cite{Linfoot:55, Sullivan:98, Ashok:03}, with no need for a processing algorithm to evaluate the design.	
It additionally opens several avenues for future research.
Ideal thin lenses could be replaced by a surrogate model \cite{Goossens2022} trained with the help of Zemax blackbox files to have access to a wider range of lenses.
The differentiability of the ray-tracing simulator can be exploited for optimizing coded apertures, potentially leading to designs that maximize information capture or minimize reconstruction error.
Dynamic mask designs could also be achieved, accurately adapting to a scene in order to process several coded acquisitions and reach better information acquisition.
Finally, this work paves the way for end-to-end optimization of CASSI systems, integrating both optical design and computational algorithms to achieve optimal performance in co-design computational imaging.

\section*{Acknowledgments}
This project has received financial support from the CNRS through the MITI interdisciplinary programs.

\bibliographystyle{IEEEtran}
\bibliography{sample}


\section{Biography Section}
\begin{IEEEbiography}[{\includegraphics[width=1in,height=1.25in,clip,keepaspectratio]{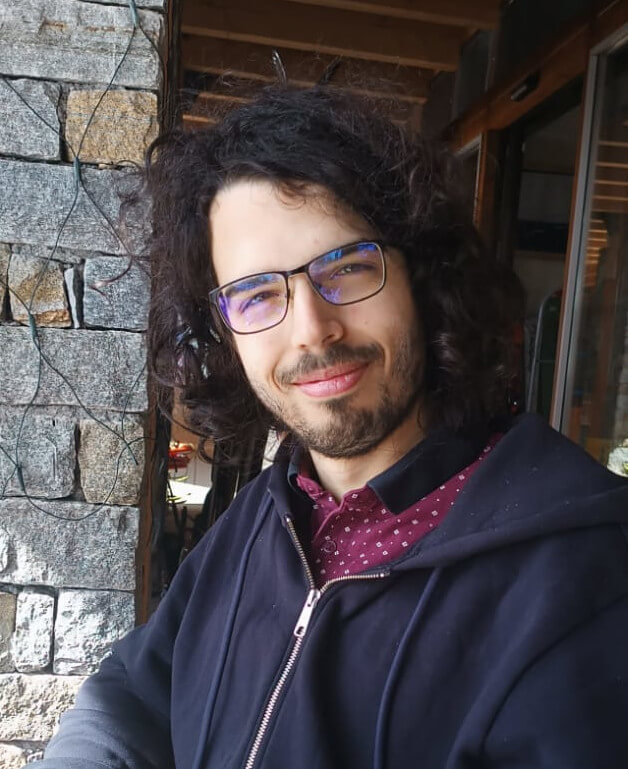}}]{Léo Paillet} received his MSc and engineering degree from École Nationale Supérieure de Techniques Avancées Paris (ENSTA Paris) and Institut Polytechnique de Paris, Palaiseau, France in 2022. He is currently pursuing his PhD degree at LAAS-CNRS and IRAP, Toulouse, France. His research interests include systems modelization and optimization, as well as computer vision through learning.\end{IEEEbiography}

\begin{IEEEbiography}[{\includegraphics[width=1in,height=1.25in,clip,keepaspectratio]{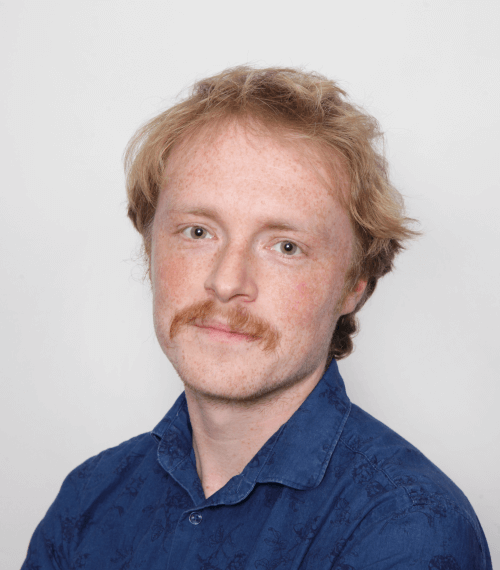}}]{Antoine Rouxel} received the Engineering degree from the Institut d'Optique Graduate School, Palaiseau, France, in 2018, and the Ph.D. degree in photonics and optoelectronic systems from LAAS--CNRS, Toulouse, France, in~2022. His research interests include the design of hybrid instruments that tightly integrate optics with computational algorithms, particularly coded‑aperture hyperspectral imagers and holographic devices. He is currently a Postdoctoral Researcher with the Department of Automation, Tsinghua University, Beijing, China, where he is developing a mesoscopic imaging and stimulation platform for neuroscience applications.
\end{IEEEbiography}

\begin{IEEEbiography}[{\includegraphics[width=1in,height=1.25in,clip,keepaspectratio]{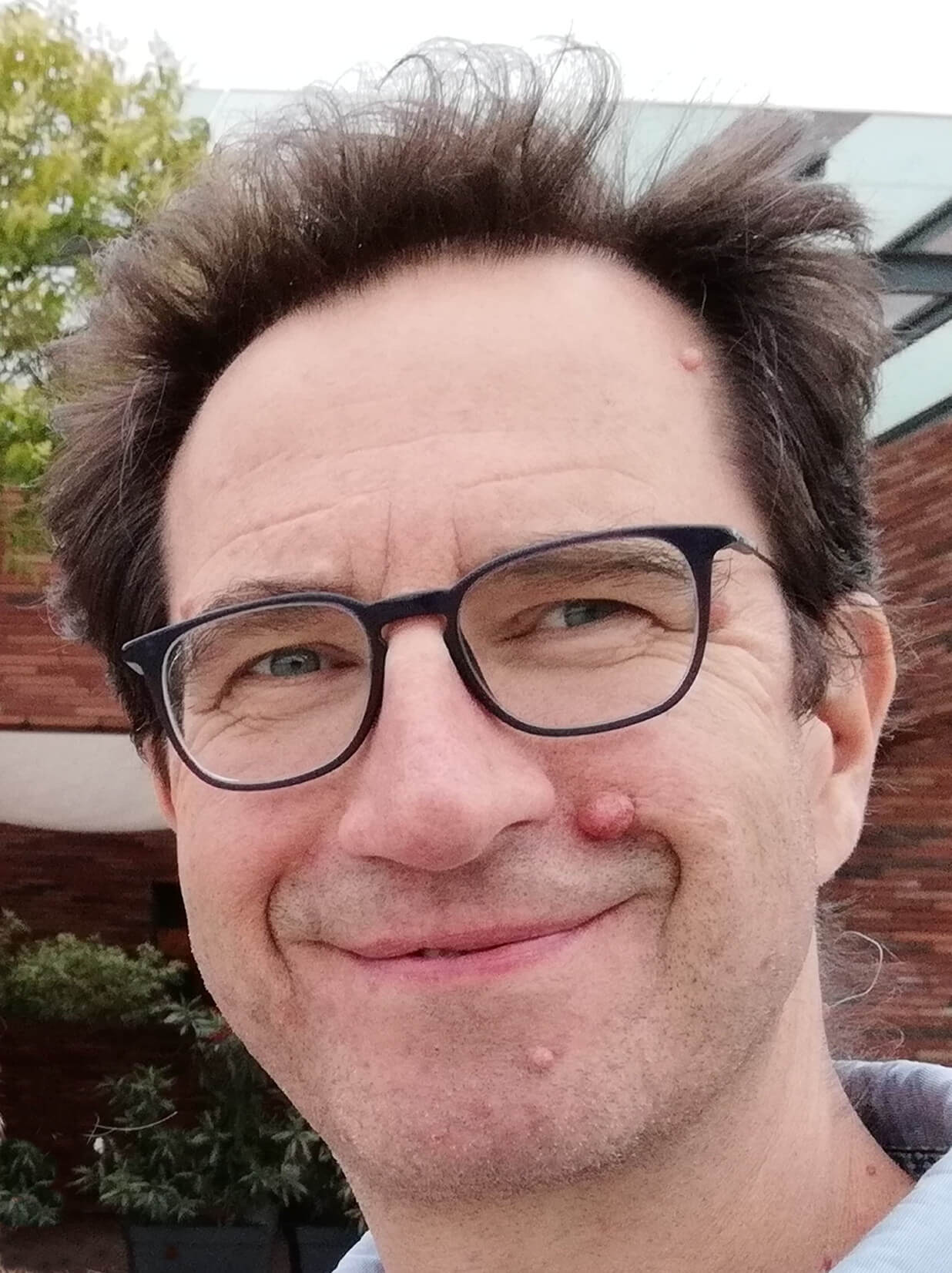}}]{Hervé Carfantan} Hervé Carfantan was born in France in 1968. He received the Engineering degree in control and signal processing from École Supérieure d’Informatique Électronique Automatique, Paris, France, in 1992, and the M.S. degree in control and signal processing and Ph.D. degree in physics from the Université de Paris-Sud, Orsay, France, in 1992 and 1996, respectively. 
He is currently a Professor with Université de Toulouse, France, and a Researcher with the Signal, Image en Sciences de l’Univers Team, Institut de Recherche en Astrophysique et Planétologie, Toulouse, which is part of the Université de Toulouse, the Centre National de la Recherche Scientifique (CNRS) and the Centre National d'Études Spatiales (CNES). 
His research interests include inverse problems, estimation, spectral analysis (in particular, the case of irregularly sampled data), sparse approximation, optimization and hyperspectral imaging. His application fields essentially concern astronomical and remote sensing data analysis.
\end{IEEEbiography}
\begin{IEEEbiography}[{\includegraphics[width=1in,height=1.25in,clip,keepaspectratio]{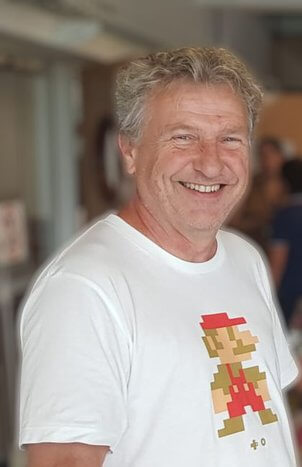}}]{Simon Lacroix} passed away in May 2025. He was a senior researcher at the Laboratory for Analysis and Architecture of Systems (LAAS), part of the Centre National de la Recherche Scientifique (CNRS) in Toulouse, France, where he led research activities in field robotics. He played a pioneering role in planetary robotics during the 1990s and initiated the laboratory’s aerial robotics efforts in the early 2000s. His work focused on the deployment of heterogeneous teams of autonomous robots for exploration, surveillance, and intervention missions. His research interests included perception and navigation for aerial and terrestrial robots—such as environment modeling, localization, perception control, and autonomous navigation strategies—and evolved toward the decisional processes underpinning multirobot cooperation, as well as active perception particularly through compressed hyperspectral imaging.
\end{IEEEbiography}
\begin{IEEEbiographynophoto}{Antoine Monmayrant} received a Ph.D. in molecular and atomic physics from the Université Paul Sabatier, Toulouse, France, in 2005.
	From 2005 to 2006, he was a Postdoctoral Researcher with the Ultrafast Group in Oxford, U.K.
	He then joined the PHOTONICS Group in the Laboratory for Analysis and Architecture of Systems (LAAS-CNRS),  Toulouse, France, where he works on photonic crystal lasers, photonic structures, guided-mode resonant filters and coded hyperspectral imaging. Photograph not available at the time of publication.
\end{IEEEbiographynophoto}
\vfill

\addtocounter{figure}{-1}
\refstepcounter{figure}\label{LASTFIGURE}
\addtocounter{table}{-1}
\refstepcounter{table}\label{LASTTABLE}
\end{document}


\title{The Marginal Importance of Known Distortions and Alignment in CASSI systems\\
--Supplementary material--}
\author{Léo Paillet, Antoine Rouxel, Hervé Carfantan, Simon Lacroix and Antoine Monmayrant
}

%


%
%
\maketitle

This supplementary material provides:
\begin{itemize}
\item More comparisons between the spot diagrams from our implementation and Zemax.
\item Details on some characteristics of the considered systems, and acquisitions rendered with them.
  \item Further results on the reconstruction quality across the four considered systems with the different reconstruction algorithms.
\end{itemize}

\section{Point spread functions comparison with Zemax}

Figures \ref{fig:psf_single}, \ref{fig:psf_single_mis}, \ref{fig:psf_amici} and \ref{fig:psf_amici_mis} show the spot diagrams estimated with Zemax and with our implementation of dO~\cite{dO}, for the (SP), (mSP), (SP), (mSP) configurations respectively.

As stated in the main document, the (mSP) configuration is the one that yields the highest RMS differences between both implementations (up to 1.2\,µm, well below the considered pixel size).
The RMS differences are smaller in the three other configurations, especially for the Amici configurations (AP) and (mAP).

Each figure shows the two spot diagrams for the three wavelengths of interest: 450\,nm, 520\,nm, 650\,nm, at four different positions in the field of view, denoted by the red dot on the bottom left of each figure.
In the figures, the dotted black line represents the RMS radius centered on the centroid of the spot diagrams.
Note the pixel size scales are varying, because the size of the spot diagrams depends on the considered configuration, wavelength and position in the field of view.

\newpage
\begin{figure}[!hbt]
	\centering
\includegraphics[width=0.85\columnwidth]{./supp_images/psf_single_full.pdf}
	\caption{Spot diagrams obtained with the (SP) configuration at four positions in the field of view and for three wavelengths. Top rows show the spot diagrams obtained with dO, bottom rows show the spot diagrams obtained with Zemax.}
	\label{fig:psf_single}
\end{figure}

\newpage
\begin{figure}[!hbt]
	\centering
\includegraphics[width=0.85\columnwidth]{./supp_images/psf_single_misaligned_full.pdf}
	\caption{Spot diagrams obtained with the (mSP) configuration at four positions in the field of view and for three wavelengths. Top rows show the spot diagrams obtained with dO, bottom rows show the spot diagrams obtained with Zemax.}
	\label{fig:psf_single_mis}
\end{figure}

\newpage
\begin{figure}[!hbt]
	\centering
\includegraphics[width=0.85\columnwidth]{./supp_images/psf_amici_full.pdf}
	\caption{Spot diagrams obtained with the (AP) configuration at four positions in the field of view and for three wavelengths. Top rows show the spot diagrams obtained with dO, bottom rows show the spot diagrams obtained with Zemax.}
	\label{fig:psf_amici}
\end{figure}

\newpage
\begin{figure}[!hbt]
	\centering
\includegraphics[width=0.85\columnwidth]{./supp_images/psf_amici_misaligned_full.pdf}
	\caption{Spot diagrams obtained with the (mAP) configuration at four positions in the field of view and for three wavelengths. Top rows show the spot diagrams obtained with dO, bottom rows show the spot diagrams obtained with Zemax.}
	\label{fig:psf_amici_mis}
\end{figure}

\newpage
\section{Configurations Characteristics}
\subsection{Acquisition rendering and spectral upsampling}
The large nonlinear spectral spreading of our configurations mentioned in\,\cref{main-subsubsec:rendering} requires a spectral upsampling of the scenes at the rendering stage.
Indeed, as our system provides a large spectral spreading (around 830\,µm) compared to its diffraction limited spot size (between 5 and 8\,µm) and its distortions (that is the extension of the spot diagrams), its spectral resolution is good enough to observe the discreteness of the CAVE and KAIST datasets.
Without upsampling, the rendered acquisitions would exhibit distinct bands corresponding to the 28 spectral bands of the datasets, with dark stripes in between.
Those stripes wouldn't appear if we were to acquire a real scene, since real spectra are continuous.
We thus decided to upsample the scenes in order to accurately render realistic acquisitions.
\begin{figure}[!h]
	\centering
	\includegraphics[width=0.75\columnwidth]{./supp_images_round1/fig6_diff_sampling.pdf}
	\caption{Spectra extracted from rendered single-slit acquisition of a scene upsampled to 28, 56, 84, 112, 224 spectral bands with the (SP) configuration compared to ground truth spectra retrieved from the HSS. The spectra of the points observed through the slit is horizontally spread on the detector.
		The ground truth spectra correspond to the HSS data spatially dispersed and convoluted by the approximated PSFs and Airy disks of the system.
		Top left - RGB image of the scene and location of the slit (dashed line).
		For each factor: Rendered acquisition with the (SP) configuration. Acquired spectra (solid lines) in the 3 regions with constant color (green, orange and gray) compared to their ground truth counterparts (dashed lines).}
	\label{fig:diff_sampling}
\end{figure}
This is illustrated in Figure\,\ref{fig:diff_sampling}, showing the rendered acquisition on the detector and retrieved spectra for a single-slit mask with upsampling varying from 28 to 224 bands.

As one can see, non-physical stripes are visible on the acquisitions for the HS scene upsampled up to 84 bands.
These stripes cause large oscillations on the retrieved spectra along a given horizontal line of the detector.
However, for the scene upsampled to 112 and 224 bands, the rendered acquisitions are smooth (as expected) and the retrieved spectra are close to the ground truth.

\subsection{Spreading non-linearity}

As shown in Figure~\ref{main-fig:dispersion_curves} of the manuscript, both configurations (AP) and (SP) exhibit non-linear spectro-spatial spreading.
We didn't focus on minimizing (AP)'s non-linearity because we only wanted comparable configurations, and (SP) already exhibits a strong non-linear spreading.
This is investigated in Figure~\ref{fig:nonlinear}, where we can see in subfigure (b) the difference of both configurations to a fitted linear spreading.
\begin{figure}[!h]
	\centering
	\includegraphics[width=0.45\columnwidth]{./supp_images_round1/dispersion_curves_w_diff_to_lin.pdf}
	\caption{Non-linear spatial dispersion.
		(a) Comparison of the spatial dispersion at the detector (spreading) for (AP) (blue) and (SP) (orange).
		(b) Difference between (AP) (blue) and (SP) (orange) and their respective fitted linear dispersion curve.
		The dots correspond to 28 considered wavelengths over our spectral range.
	}
	\label{fig:nonlinear}
\end{figure}
Both configurations exhibit distinct non-linear spreading with a maximum  of more than 79\,µm (7.9 pixels) for (SP).
We can see that (AP) has a slightly stronger non-linear spreading with differences to the linear spreading of up to 110\,µm (11 pixels).
Therefore, the non-linearity has to be accounted for in the mapping function described in section~\ref{main-sec:mapping}.

\subsection{Distortions}
\label{sec:distortions}

The distortions for the four configurations are illustrated in Figure\,\ref{fig:spot_diagram} by tracing rays from 9 positions in the field of view, at 3 different wavelengths (minimum, central and maximum), and recording their position of arrival on the detector (color dots).
Additionally, three vertical lines at the right, center and left of the field of view and three horizontal lines at the bottom, center and top of the field of view are also imaged (colored lines -- with 20 points per line).
\begin{figure}[!hbt]
	\centering
\includegraphics[width=0.75\columnwidth]{./supp_images/spot_diagram.pdf}
	\caption{Illustration of the distortions for the four CASSI configurations.}
	\label{fig:spot_diagram}
\end{figure}

From these points and lines, we can evaluate both keystone and smile distortions.

Keystone distortion corresponds to a vertical magnification that vary with both the wavelength and the lateral position, affecting how an horizontal line is imaged.
Its image is no longer horizontal and its vertical position depend on the wavelength.
To estimate the keystone, for each horizontal line we measure the vertical extent of its image (i.e.\ max to min values of the vertical position) across the three wavelength.

Smile distortion corresponds to the bowing observed on the image of a vertical line in the scene.
It varies with both wavelength and lateral position.
We estimate the smile distortion by measuring the lateral extent of each bowed lines and taking the maximum over both position and wavelength.

For perfectly aligned systems, the system with Amici prism assembly (AP) does not exhibit significant smile distortion (3\,µm thus below one 10-µm-wide pixel) and a keystone that is barely measurable (11\,µm or around 1.1\,pixel) whereas the system based on a single prism (SP) suffers from a significant smile distortion (94\,µm or 9.4\,pixel) and a larger keystone (25\,µm or 2.5\,pixel).

For the purposely misaligned configurations the impact of misalignment is limited for the system with Amici prism assembly (mAP) with a smile distortion of 17\,µm (1.7\,pixel) and a keystone of 31\,µm (3.1\,pixel).
The configuration with a single prism (mSP) is much more impacted with estimated keystone of 318\,µm (31.8\,pixel) and smile of 280\,µm (28\,pixel).
However, as one can see in Figure\,\ref{fig:spot_diagram}, for (mSP), these large distortions go beyond the usual definition of keystone and smile distortions.

\subsection{Acquisitions}

Figure \ref{fig:acquisition} shows two typical acquisitions with the four configurations, for a given scene from both CAVE~\cite{CAVE} and KAIST dataset~\cite{Choi:17} (see \Cref{fig:acquisition}).
It illustrates the impact on the rendered images of the distortions and misalignment for each configuration.

\begin{figure}[!hbt]
	\centering
	\includegraphics[width=0.7\columnwidth]{./supp_images/acquisitions.pdf}
	\caption{Example coded acquisition for one hyperspectral scene from CAVE (a) and one from KAIST (b).}
	\label{fig:acquisition}
\end{figure}

\newpage

\section{Reconstruction results}
\label{sec:recons}

The following five tables \Cref{tab:results_dgsmp,tab:results_mst,tab:results_dauhst,tab:results_duf,tab:results_padut} show the reconstruction results with the four configurations under study of the five state-of-the-art reconstruction algorithms.
The results are detailed for each scene, the average and the standard deviation over the scenes are reported in the rightmost column.
The four considered metrics are the same as the ones in the main document: Root Mean Square Error (RMSE), Peak Signal-to-Noise Ratio (PSNR), Structural Similarity Index Measure (SSIM), Spectral Angle Mapper (SAM).
For the five algorithms and all scenes, all metrics are similar for the four considered configurations, confirming the results observed on the average over the scenes.

\begin{table}[!hbtp]
	\centering
	\begin{tabular}{>{\bfseries} c c c c c c c c c c c c c}
		\toprule
		\rowcolor{gray!20}
		& & S1 & S2 & S3 & S4 & S5 & S6 & S7 & S8 & S9 & S10 & Avg \\
		\midrule
		\multirow{4}{*}{\begin{tabular}{c}RMSE $\downarrow$\\($\times 10^{-3}$)\end{tabular}}& (AP)& 26.6& 27.4& 25.9& 14.6& 40.2& 28.7& 33.2& 40.6& 38.2& 37.3& 31.3 $\pm$ 7.8 \\
		& (SP)& 27.8& 29.1& 29.3& 13.8& 43.4& 29.5& 36.1& 37.0& 42.5& 39.0& 32.8 $\pm$ 8.4 \\
		& (mAP)& 27.4& 28.6& 29.6& 15.1& 38.2& 28.8& 33.5& 39.5& 39.3& 37.8& 31.8 $\pm$ 7.2 \\
		& (mSP)& 28.7& 31.0& 29.3& 16.8& 42.1& 31.9& 37.1& 42.0& 46.8& 38.4& 34.4 $\pm$ 8.3 \\
		\midrule
		\multirow{4}{*}{PSNR $\uparrow$}& (AP)& 31.6& 31.3& 31.9& 37.8& 28.3& 30.8& 29.6& 28.0& 28.5& 28.6& 30.6 $\pm$ 2.76 \\
		& (SP)& 31.2& 30.7& 30.9& 38.7& 27.5& 30.6& 28.9& 28.6& 27.6& 28.2& 30.3 $\pm$ 3.12 \\
		& (mAP)& 31.3& 30.9& 30.7& 37.8& 28.6& 30.8& 29.6& 28.1& 28.3& 28.5& 30.5 $\pm$ 2.72 \\
		& (mSP)& 30.9& 30.2& 30.8& 37.9& 27.7& 29.9& 28.7& 27.5& 26.7& 28.3& 29.9 $\pm$ 3.01 \\
		\midrule
		\multirow{4}{*}{\begin{tabular}{c}SSIM $\uparrow$\\$[0-1]$\end{tabular}}& (AP)& 0.903& 0.873& 0.905& 0.952& 0.889& 0.914& 0.884& 0.888& 0.870& 0.845& 0.892 $\pm$ 0.028 \\
		& (SP)& 0.895& 0.858& 0.890& 0.959& 0.871& 0.908& 0.874& 0.892& 0.846& 0.834& 0.883 $\pm$ 0.034 \\
		& (mAP)& 0.898& 0.865& 0.888& 0.951& 0.883& 0.910& 0.878& 0.886& 0.859& 0.842& 0.886 $\pm$ 0.029 \\
		& (mSP)& 0.890& 0.851& 0.892& 0.955& 0.875& 0.903& 0.867& 0.877& 0.843& 0.833& 0.879 $\pm$ 0.033 \\
		\midrule
		\multirow{4}{*}{\begin{tabular}{c}SAM $\downarrow$\\$[0-1]$\end{tabular}}& (AP)& 0.057& 0.066& 0.052& 0.052& 0.048& 0.060& 0.052& 0.078& 0.061& 0.055& 0.058 $\pm$ 0.008 \\
		& (SP)& 0.058& 0.072& 0.058& 0.052& 0.057& 0.065& 0.056& 0.077& 0.072& 0.058& 0.062 $\pm$ 0.008 \\
		& (mAP)& 0.054& 0.072& 0.060& 0.052& 0.050& 0.065& 0.053& 0.074& 0.060& 0.055& 0.059 $\pm$ 0.008 \\
		& (mSP)& 0.063& 0.076& 0.058& 0.059& 0.057& 0.069& 0.058& 0.083& 0.072& 0.059& 0.065 $\pm$ 0.009 \\
		\bottomrule
	\end{tabular}
	\caption{RMSE, PSNR (in dB), SSIM and SAM for DGSMP for each optical system.}
	\label{tab:results_dgsmp}
\end{table}

\begin{table}[!hbtp]
	\centering
	\begin{tabular}{>{\bfseries} c c c c c c c c c c c c c}
		\toprule
		\rowcolor{gray!20}
		& & S1 & S2 & S3 & S4 & S5 & S6 & S7 & S8 & S9 & S10 & Avg \\
		\midrule
		\multirow{4}{*}{\begin{tabular}{c}RMSE $\downarrow$\\($\times 10^{-3}$)\end{tabular}}& (AP)& 24.4& 25.2& 23.9& 12.3& 32.0& 25.8& 32.2& 33.6& 31.0& 32.4& 27.3 $\pm$ 6.1 \\
		& (SP)& 23.7& 24.5& 21.7& 11.8& 32.2& 25.7& 31.2& 31.8& 29.9& 32.4& 26.5 $\pm$ 6.2 \\
		& (mAP)& 23.0& 25.1& 23.2& 12.7& 31.5& 24.2& 31.8& 32.2& 30.3& 30.8& 26.5 $\pm$ 5.8 \\
		& (mSP)& 23.2& 24.5& 23.0& 12.5& 31.1& 26.0& 31.6& 33.6& 31.5& 33.0& 27.0 $\pm$ 6.2 \\
		\midrule
		\multirow{4}{*}{PSNR $\uparrow$}& (AP)& 32.5& 32.1& 32.5& 40.1& 30.3& 31.8& 29.9& 29.7& 30.5& 29.8& 31.9 $\pm$ 2.92 \\
		& (SP)& 32.7& 32.3& 33.3& 40.5& 30.2& 31.8& 30.2& 30.0& 30.7& 29.8& 32.2 $\pm$ 3.02 \\
		& (mAP)& 33.0& 32.1& 32.7& 39.9& 30.5& 32.3& 30.0& 30.0& 30.6& 30.3& 32.1 $\pm$ 2.81 \\
		& (mSP)& 32.8& 32.3& 32.8& 39.8& 30.5& 31.7& 30.1& 29.6& 30.4& 29.7& 32.0 $\pm$ 2.88 \\
		\midrule
		\multirow{4}{*}{\begin{tabular}{c}SSIM $\uparrow$\\$[0-1]$\end{tabular}}& (AP)& 0.917& 0.885& 0.923& 0.965& 0.916& 0.925& 0.878& 0.907& 0.901& 0.878& 0.910 $\pm$ 0.025 \\
		& (SP)& 0.922& 0.893& 0.932& 0.969& 0.916& 0.924& 0.890& 0.913& 0.901& 0.879& 0.914 $\pm$ 0.024 \\
		& (mAP)& 0.922& 0.890& 0.924& 0.960& 0.916& 0.928& 0.883& 0.912& 0.899& 0.884& 0.912 $\pm$ 0.023 \\
		& (mSP)& 0.920& 0.890& 0.932& 0.967& 0.915& 0.924& 0.886& 0.904& 0.897& 0.871& 0.911 $\pm$ 0.026 \\
		\midrule
		\multirow{4}{*}{\begin{tabular}{c}SAM $\downarrow$\\$[0-1]$\end{tabular}}& (AP)& 0.053& 0.061& 0.045& 0.067& 0.044& 0.058& 0.049& 0.071& 0.050& 0.055& 0.055 $\pm$ 0.008 \\
		& (SP)& 0.050& 0.059& 0.043& 0.056& 0.041& 0.056& 0.047& 0.067& 0.049& 0.055& 0.052 $\pm$ 0.008 \\
		& (mAP)& 0.051& 0.060& 0.044& 0.061& 0.042& 0.061& 0.048& 0.072& 0.051& 0.057& 0.055 $\pm$ 0.009 \\
		& (mSP)& 0.049& 0.058& 0.044& 0.061& 0.044& 0.058& 0.049& 0.072& 0.052& 0.062& 0.055 $\pm$ 0.008 \\
		\bottomrule
	\end{tabular}
	\caption{RMSE, PSNR (in dB), SSIM and SAM for MST for each optical system.}
	\label{tab:results_mst}
\end{table}

\begin{table}[!hbtp]
	\centering
	\begin{tabular}{>{\bfseries} c c c c c c c c c c c c c}
		\toprule
		\rowcolor{gray!20}
		& & S1 & S2 & S3 & S4 & S5 & S6 & S7 & S8 & S9 & S10 & Avg \\
		\midrule
		\multirow{4}{*}{\begin{tabular}{c}RMSE $\downarrow$\\($\times 10^{-3}$)\end{tabular}}& (AP)& 19.4& 17.7& 17.6& 12.0& 25.8& 19.0& 21.3& 25.1& 23.2& 26.3& 20.7 $\pm$ 4.3 \\
		& (SP)& 18.1& 16.7& 16.7& 10.3& 24.4& 19.3& 20.6& 24.7& 19.9& 25.7& 19.6 $\pm$ 4.4 \\
		& (mAP)& 18.5& 16.7& 17.3& 10.3& 24.8& 18.4& 20.4& 24.0& 19.9& 24.9& 19.5 $\pm$ 4.2 \\
		& (mSP)& 18.6& 18.0& 17.4& 10.1& 25.5& 19.4& 21.6& 24.8& 21.6& 26.2& 20.3 $\pm$ 4.6 \\
		\midrule
		\multirow{4}{*}{PSNR $\uparrow$}& (AP)& 34.7& 35.3& 35.4& 40.8& 32.5& 34.5& 33.4& 32.2& 33.2& 31.8& 34.4 $\pm$ 2.44 \\
		& (SP)& 35.1& 35.7& 35.7& 41.2& 32.8& 34.4& 33.7& 32.3& 34.3& 32.0& 34.7 $\pm$ 2.51 \\
		& (mAP)& 35.0& 35.8& 35.5& 41.7& 32.8& 34.8& 33.8& 32.6& 34.4& 32.2& 34.9 $\pm$ 2.57 \\
		& (mSP)& 34.8& 35.0& 35.3& 41.4& 32.4& 34.3& 33.3& 32.2& 33.6& 31.8& 34.4 $\pm$ 2.61 \\
		\midrule
		\multirow{4}{*}{\begin{tabular}{c}SSIM $\uparrow$\\$[0-1]$\end{tabular}}& (AP)& 0.940& 0.935& 0.951& 0.973& 0.942& 0.951& 0.933& 0.939& 0.937& 0.917& 0.942 $\pm$ 0.014 \\
		& (SP)& 0.945& 0.934& 0.952& 0.976& 0.944& 0.951& 0.937& 0.941& 0.946& 0.921& 0.945 $\pm$ 0.014 \\
		& (mAP)& 0.943& 0.940& 0.949& 0.979& 0.945& 0.953& 0.938& 0.938& 0.940& 0.923& 0.945 $\pm$ 0.014 \\
		& (mSP)& 0.940& 0.925& 0.947& 0.973& 0.938& 0.944& 0.929& 0.936& 0.940& 0.914& 0.939 $\pm$ 0.015 \\
		\midrule
		\multirow{4}{*}{\begin{tabular}{c}SAM $\downarrow$\\$[0-1]$\end{tabular}}& (AP)& 0.047& 0.053& 0.038& 0.055& 0.040& 0.054& 0.036& 0.062& 0.045& 0.054& 0.048 $\pm$ 0.008 \\
		& (SP)& 0.044& 0.054& 0.037& 0.056& 0.039& 0.055& 0.034& 0.060& 0.043& 0.059& 0.048 $\pm$ 0.009 \\
		& (mAP)& 0.044& 0.053& 0.036& 0.053& 0.040& 0.055& 0.034& 0.064& 0.041& 0.053& 0.047 $\pm$ 0.009 \\
		& (mSP)& 0.048& 0.056& 0.038& 0.057& 0.042& 0.058& 0.036& 0.062& 0.044& 0.056& 0.050 $\pm$ 0.009 \\
		\bottomrule
	\end{tabular}
	\caption{RMSE, PSNR (in dB), SSIM and SAM for DAUHST for each optical system.}
	\label{tab:results_dauhst}
\end{table}

\begin{table}[!hbtp]
	\centering
	\begin{tabular}{>{\bfseries} c c c c c c c c c c c c c}
		\toprule
		\rowcolor{gray!20}
		& & S1 & S2 & S3 & S4 & S5 & S6 & S7 & S8 & S9 & S10 & Avg \\
		\midrule
		\multirow{4}{*}{\begin{tabular}{c}RMSE $\downarrow$\\($\times 10^{-3}$)\end{tabular}}& (AP)& 20.4& 19.2& 20.1& 12.3& 28.6& 21.6& 24.2& 28.5& 23.3& 27.9& 22.6 $\pm$ 4.8 \\
		& (SP)& 22.4& 20.9& 20.4& 13.8& 29.6& 22.5& 27.3& 29.1& 25.3& 30.2& 24.1 $\pm$ 4.9 \\
		& (mAP)& 21.5& 20.9& 19.8& 12.5& 29.7& 21.5& 25.8& 30.3& 26.4& 29.8& 23.8 $\pm$ 5.3 \\
		& (mSP)& 21.7& 20.9& 20.4& 11.2& 28.6& 22.7& 26.3& 29.9& 25.9& 29.2& 23.7 $\pm$ 5.3 \\
		\midrule
		\multirow{4}{*}{PSNR $\uparrow$}& (AP)& 34.0& 34.5& 34.1& 40.3& 31.4& 33.4& 32.3& 31.1& 32.8& 31.2& 33.5 $\pm$ 2.55 \\
		& (SP)& 33.1& 33.7& 33.9& 39.6& 30.9& 33.0& 31.4& 30.8& 32.1& 30.5& 32.9 $\pm$ 2.53 \\
		& (mAP)& 33.5& 33.7& 34.1& 39.9& 30.9& 33.4& 31.8& 30.5& 31.8& 30.7& 33.0 $\pm$ 2.62 \\
		& (mSP)& 33.4& 33.6& 33.8& 40.5& 31.1& 32.9& 31.6& 30.5& 31.9& 30.8& 33.0 $\pm$ 2.76 \\
		\midrule
		\multirow{4}{*}{\begin{tabular}{c}SSIM $\uparrow$\\$[0-1]$\end{tabular}}& (AP)& 0.936& 0.921& 0.933& 0.969& 0.929& 0.940& 0.918& 0.925& 0.929& 0.905& 0.930 $\pm$ 0.016 \\
		& (SP)& 0.926& 0.909& 0.930& 0.968& 0.921& 0.935& 0.903& 0.919& 0.916& 0.893& 0.922 $\pm$ 0.019 \\
		& (mAP)& 0.931& 0.914& 0.934& 0.967& 0.925& 0.940& 0.912& 0.919& 0.919& 0.898& 0.926 $\pm$ 0.018 \\
		& (mSP)& 0.928& 0.914& 0.933& 0.971& 0.927& 0.937& 0.910& 0.919& 0.920& 0.898& 0.926 $\pm$ 0.019 \\
		\midrule
		\multirow{4}{*}{\begin{tabular}{c}SAM $\downarrow$\\$[0-1]$\end{tabular}}& (AP)& 0.048& 0.059& 0.041& 0.058& 0.048& 0.058& 0.039& 0.075& 0.049& 0.058& 0.053 $\pm$ 0.010 \\
		& (SP)& 0.051& 0.061& 0.045& 0.054& 0.047& 0.060& 0.045& 0.077& 0.050& 0.061& 0.055 $\pm$ 0.010 \\
		& (mAP)& 0.051& 0.058& 0.041& 0.059& 0.048& 0.058& 0.042& 0.078& 0.048& 0.056& 0.054 $\pm$ 0.010 \\
		& (mSP)& 0.051& 0.061& 0.042& 0.055& 0.046& 0.056& 0.044& 0.075& 0.048& 0.057& 0.053 $\pm$ 0.009 \\
		\bottomrule
	\end{tabular}
	\caption{RMSE, PSNR (in dB), SSIM and SAM for DUF for each optical system.}
	\label{tab:results_duf}
\end{table}

\begin{table}[!hbtp]
	\centering
	\begin{tabular}{>{\bfseries} c c c c c c c c c c c c c}
		\toprule
		\rowcolor{gray!20}
		& & S1 & S2 & S3 & S4 & S5 & S6 & S7 & S8 & S9 & S10 & Avg \\
		\midrule
		\multirow{4}{*}{\begin{tabular}{c}RMSE $\downarrow$\\($\times 10^{-3}$)\end{tabular}}& (AP)& 19.5& 18.2& 19.4& 10.3& 27.4& 19.4& 23.4& 26.0& 23.2& 27.4& 21.4 $\pm$ 5.0 \\
		& (SP)& 18.2& 17.3& 17.4& 10.0& 23.8& 18.5& 22.3& 23.0& 20.5& 24.6& 19.6 $\pm$ 4.1 \\
		& (mAP)& 19.8& 18.0& 19.0& 11.6& 27.1& 20.2& 23.2& 26.8& 23.4& 26.5& 21.6 $\pm$ 4.6 \\
		& (mSP)& 20.2& 19.0& 18.5& 11.0& 27.9& 20.8& 23.7& 26.2& 23.3& 26.9& 21.8 $\pm$ 4.8 \\
		\midrule
		\multirow{4}{*}{PSNR $\uparrow$}& (AP)& 34.7& 35.1& 34.6& 40.9& 32.1& 34.4& 32.6& 32.1& 33.2& 31.5& 34.1 $\pm$ 2.57 \\
		& (SP)& 35.1& 35.3& 35.3& 41.2& 32.9& 34.7& 33.0& 32.8& 34.0& 32.3& 34.7 $\pm$ 2.43 \\
		& (mAP)& 34.5& 35.1& 34.7& 40.7& 32.0& 34.0& 32.7& 31.7& 32.9& 31.7& 34.0 $\pm$ 2.54 \\
		& (mSP)& 34.2& 34.5& 34.7& 40.5& 31.7& 33.7& 32.5& 31.7& 32.9& 31.5& 33.8 $\pm$ 2.51 \\
		\midrule
		\multirow{4}{*}{\begin{tabular}{c}SSIM $\uparrow$\\$[0-1]$\end{tabular}}& (AP)& 0.941& 0.931& 0.943& 0.968& 0.936& 0.950& 0.922& 0.936& 0.930& 0.915& 0.937 $\pm$ 0.014 \\
		& (SP)& 0.944& 0.934& 0.949& 0.972& 0.944& 0.951& 0.926& 0.941& 0.940& 0.925& 0.943 $\pm$ 0.013 \\
		& (mAP)& 0.940& 0.930& 0.944& 0.967& 0.935& 0.947& 0.923& 0.932& 0.926& 0.916& 0.936 $\pm$ 0.014 \\
		& (mSP)& 0.937& 0.924& 0.944& 0.967& 0.933& 0.945& 0.923& 0.931& 0.932& 0.913& 0.935 $\pm$ 0.014 \\
		\midrule
		\multirow{4}{*}{\begin{tabular}{c}SAM $\downarrow$\\$[0-1]$\end{tabular}}& (AP)& 0.044& 0.051& 0.036& 0.056& 0.043& 0.054& 0.036& 0.062& 0.042& 0.053& 0.048 $\pm$ 0.008 \\
		& (SP)& 0.045& 0.052& 0.036& 0.057& 0.039& 0.056& 0.036& 0.061& 0.042& 0.054& 0.048 $\pm$ 0.009 \\
		& (mAP)& 0.045& 0.053& 0.036& 0.058& 0.041& 0.055& 0.037& 0.066& 0.043& 0.052& 0.049 $\pm$ 0.009 \\
		& (mSP)& 0.047& 0.053& 0.036& 0.056& 0.042& 0.056& 0.037& 0.062& 0.043& 0.051& 0.048 $\pm$ 0.008 \\
		\bottomrule
	\end{tabular}
	\caption{RMSE, PSNR (in dB), SSIM and SAM for PADUT for each optical system.}
	\label{tab:results_padut}
\end{table}

\newpage 

\subsection{RMSE}

The \textit{Root Mean Square Error} (RMSE) describes the mean error between two quantities.
For objects $X$ and $Y$ with $n$ elements, it is defined as:
\begin{equation}
	\text{RMSE}(X,Y) = \sqrt{\frac{1}{n}\sum_{i=1}^n (X_i - Y_i)^2}
\end{equation}
For hyperspectral images, we compute the RMSE on the whole hyperspectral cubes, so $n = HWK.$

\subsection{PSNR}

The \textit{Peak signal-to-noise ratio} (PSNR) is defined as the ratio between the maximum intensity of an object and the difference between its noisy (reconstructed in our case) and noise-free representations.
For objects $X$ and $Y$ with $n$ elements ($n = HWK$ for hyperspectral cubes) with values within the $[0, 255]$ range, it is defined in dB as:
\begin{equation}
	\text{PSNR}(X,Y) = 20\log_{10}\left ( \frac{255}{\text{RMSE}(X,Y)}\right )
\end{equation}

\subsection{SSIM}

The \textit{Structural Similarity Index Measure} (SSIM) [4] is based on the perceived similarity of two images.
It uses a $N \times N$ window to compute a similarity value on all such windows.
The values are then averaged over all windows.
For a given windows $x,y$ in a given spectral plane from respectively the reconstructed cube and the scene, the SSIM is defined as:
\begin{equation}
	\text{SSIM}(x, y) = \frac{(2\mu_x \mu_y + c_1)(2\sigma_{xy} + c_2)}{(\mu_x^2 + \mu_y^2 + c_1)(\sigma_x^2 + \sigma_y^2 + c_2)}
\end{equation}
with $\mu_x, \mu_y$ the sample mean intensity in windows $x,y$; $\sigma_x^2, \sigma_y^2$ the sample variance; $\sigma_{xy}$ the sample covariance of $x$ and $y$ and $c_1, c_2$ constants.
For hyperspectral images, we compute the similarity between each spectral plane (out of 28) of the scene and reconstructed hyperspectral cube using windows of size $11 \times 11$.
Afterwards, the $28$ obtained SSIM values are averaged to reach a single metric in the rang [0,1], where~1 corresponds to identical hyperspectral images.\\

\subsection{SAM}

The \textit{Spectral Angle Mapper} (SAM) [5] is a measure of similarity between two spectra which is insensitive to illumination.
It evaluates such similarity by calculating the angle between the spectra, with small angles indicating higher similarity.
For a given pair of spectra $x, y$ of size $K$ ($K=28$ in our case), the SAM is calculated as:
\begin{equation}
	\text{SAM}(x, y) = \frac{1}{\pi}\cos^{-1} \left ( \frac{\sum_{k=1}^K x_ky_k}{||x||_2 ||y||_2} \right )
\end{equation}
The usual definition of the SAM is given in degrees or radians.
We have decided to normalize it between 0 and 1, with~0 corresponding to identical spectra, so that SAM and SSIM are in the same range of values.		
For hyperspectral images, we compute the average of the SAMs between all the pairs of corresponding spectra from the reconstructed cube and the scene.

\clearpage
\bibliographystyle{IEEEtran}
\bibliography{sample}